\documentclass[12pt]{iopart}
\usepackage{iopams}
\usepackage{amssymb}
\usepackage{graphicx}
\usepackage{dsfont}
\usepackage{bm}
\newcommand{\Op}[1]{{{\mathrm{\hat{#1}}}}}

\newcommand{\beq}{\begin{equation}}
\newcommand{\eeq}{\end{equation}}

\newcommand{\beqa}{\begin{eqnarray}}
\newcommand{\eeqa}{\end{eqnarray}}
\newcommand{\beqas}{\begin{eqnarray*}}
\newcommand{\eeqas}{\end{eqnarray*}}

\begin{document}
\title{Activated and non activated dephasing demonstrated in NV center dynamics}
\author{E. Torrontegui$^{1,2}$ and R. Kosloff$^1$}

\address{$^1$ Institute of Chemistry and The Fritz Haber Research Center, The Hebrew University, Jerusalem 91904, Israel}
\address{$^2$ Departamento de Qu\'{\i}mica F\'{\i}sica, Universidad del Pa\'{\i}s Vasco - Euskal Herriko Unibertsitatea, Apdo. 644, Bilbao, Spain}
\begin{abstract}
We analyze different decoherence processes in a system coupled to a bath. Apart from the well known standard dephasing
mechanism which is temperature dependent an alternative mechanism is presented, the spin-swap dephasing 
which does not need initial bath activation and is temperature independent. 
We show that for dipolar interaction the separation of time scales between system and bath can not produce pure dephasing, the process being accompained
by dissipation.
Activated and non activated dephasing processes are demonstrated 
in a diamond nitrogen-vacancy (NV) center.
\end{abstract}  	
%
%
\section{Introduction}
{\it Decoherence} is a fundamental process where a  quantum system loses its wave properties enamelling interference.  
As a result classical behavior emerges. Decoherence has been studied at least for 50 years \cite{Fey} nevertheless, it is still ill defined \cite{Lukin, hanson, lange, gijs, nori, Knowles, nakamura}.
New quantum technologies require control on the decoherence and knowledge on the transition from quantum to classical behaviour \cite{Brune, Giulini, Arndt, Lorente}.
In the implementation of possible future quantum technologies \cite{Alber} and quantum information processing \cite{Ek, Rai} fast decoherence is destructive. 
The signature of decoherence is purity loss where purity is defined by ${\cal P}= tr \{ \Op \rho^2 \}$.

The origin of decoherence is the interaction of a system with its environment i.e., no quantum system is isolated.  The general  framework is of a mesoscopic or even macroscopic primary systems described by a hamiltonian $\Op H_S$ interacting 
through $\Op H_{SB}$ with a background environment or a bath described by $\Op H_B$, 
\beq
\Op H=\Op H_S+\Op H_B+\Op H_{SB}.
\label{H}
\eeq
The interaction generates quantum and classical correlation between the system and the environment.
One outcome is  energy transfer  between the system and bath termed {\it dissipation}.
In quantum mechanics this dissipation is accompanied by decoherence. In addition, even 
when the energy exchange is negligible, decoherence is possible. 
This scenario is termed pure decoherence or simply {\it dephasing}. Pure dephasing can be defined as loss of quantum purity
but maintaining the expectation of the energy projections. 
To quantify this loss one can employ the measure of coherence \cite{plenio}, ${\cal C}_{l_1}(\Op\rho_S) = \sum_{i\neq j}|\Op\rho_S|$ 
where we sum over the off diagonal elements of the reduced density operator of the system $\Op \rho_s=tr^{(B)}\Op\rho$ denoting
$tr^{(B)}$ the partial trace over the bath Hilbert space $\mathcal{H}_B$:

More than one scenario can generate pure  dephasing. The obvious case is when the hamiltonian $\Op H_S$ describing the system commutes with the interaction operator $\Op H_{SB}$. One can think of the environment as randomly modulating the system
hamiltonian thus generating a loss of phase during evolution. The scenario includes the case of the system energy being monitored
continuously. Typically, this modulation is produced by single interactions between bath particles, higher order bath processes can also provide relaxation mechanisms \cite{schon}.
Another scenario for pure dephasing, is the result of a separation of time scales between the system and bath. 
This will occur when there is an energy spectral gap between the system and the bath   \cite{Alicki}. 
The process can be interpreted as the  Bohr frequencies of the bath being out of range of the system Bohr frequencies
so that energy exchange between system and bath is prohibited.

Excitation of the bath can influence both mechanisms. Random modulation induced by the bath requires many of the bath modes to be excited.
As a result this mechanism of dephasing is activated and subsequently, at low temperature it can be frozen out.
The second mechanism of timescale separation is influenced by the bath excitation due to the possibility that the excitation modifies 
the spectrum, possibly changing pure dephasing  to energy dissipation.

Noteworthily, only two basic types of environment have been studied,
either a spin bath model \cite{Prok} or a 'spin-boson' oscillator bath model \cite{Fey}. Spin baths, composed of  two level systems (TLS), are geared
for the description of low energy dynamics of localized environmental modes. The difficulty in spin baths is the lack of a procedure
to obtain the parameters defining the values experimentally. At low temperature a spin bath coincides with the
spin-boson bath \cite{Prok}. Spin-boson baths are used to model a central system weakly coupled to $K$ environmental modes that
are best adapted to delocalized modes with couplings $\sim K^{-1/2}$. However, at low energies, thermodynamical variables like entropy
in almost any real physical system, are dominated by localized modes \cite{Lou} relaxing very slowly at low temperature due to system-bath couplings
independent of $K$ \cite{Prok}. The linear coupling  to spin-boson baths
is not a universal model of decoherence \cite{Alicki} and harmonic baths are not a generic quantum environment \cite{Wigner}. Nevertheless due
to the similarity between classical and quantum harmonic baths they are easy to construct because parameters can be  obtained
from classical molecular dynamics \cite{Bader} and indeed they are the starting point of many system-bath models \cite{Fey, Leg, Weiss, Makri}.
In contrast, a spin bath is universal and constitutes a universal quantum simulator. Universal quantum gates can be built from 6 spins \cite{Deutsch} manipulated 
to simulate the behavior of arbitrary
quantum systems whose dynamics are determined by local interactions \cite{lloyd}. 

Historically, the main approach towards open system dynamics is to construct equations of motion for the primary system 
where the bath is treated implicitly.
To this end, two different frameworks have been pursued: The weak coupling limit based on perturbation theory \cite{Pollard, May}, and the phenomenological dynamical semigroup formalism \cite{Gor, Lin}. 
The starting point for the perturbation theory is decomposition (\ref{H}) of the total hamiltonian and assumes weak coupling $\Op H_{SB}$.
The small parameter is the system-bath coupling. Typically, using projecting operator  
techniques \cite{May, Kubo} the equation of motion for the bath can be decoupled and solved assuming that the bath is maintained in equilibrium. 
The result is an  equation of motion  for the dynamics of the reduced density operator $\Op \rho_S(t)$ of the primary system, the so called Quantum Master Equation (QME)\cite{Breuer}. The construction leads to an integro-differetial  non-Markovian
equation. The memory effects are the price to pay for a reduced description of the primary system. 
Further reduction, assuming the bath is fast (secular approximation), leads to a Markovian equation of motion \cite{davis}.

The starting point of the dynamical semigroup formalism is the condition of complete positivity \cite{Lin2} and the Markovian assumption. The approach also provides
an equation for the dynamics of the reduced density operator $\Op \rho_S(t)$ of the primary system. 
In the so called Lindblad form \cite{Gor, Lin} the system is driven by its hamiltonian $\Op H_S$ and the interaction with the
bath is modeling by Liouville super-operators \cite{Gor}, operators acting on operators. 
The weak coupling limit can lead to the Lindblad form \cite{davis}, but the Lindblad form can be obtained in other limits
such as the singular bath limit \cite{Kossakowski} or the scattering Poissonian model \cite{lutczka}.
The Markovian assumption is equivalent to a tensor product structure of the system and bath at all times $\Op \rho = \Op \rho_S \otimes \Op \rho_B$.

Approaches to reach beyond the Markovian approximation and possibly the weak coupling limit have been considered. 
Using perturbation theory it is possible to go beyond the second
order terms like in the QME  taking higher orders \cite{Gol, Jang}. 
For exponential decaying memory it is possible to embed the dynamics in a larger space leading to coupled Markovian equations \cite{Meier}. 
Non-hermitian projection operators have also been proposed \cite{Wilkie} to include the memory
effect in the bath. 

The present paper studies pure dephasing processes based on an alternative approach which can deal with strong coupling avoiding the Markovian approximations. 
This paper is structured as follows. In sec. \ref{SH} we introduce an alternative approach, the Surrogate Hamiltonian (SH) to analyze dephasing. In addition,
we also present the spin swap process in the context of the SH. Section \ref{ho} describes a well known model of an harmonic oscillator coupled
to a bath of spins to analyze the decoherence of the harmonic oscillator. Apart from energy dissipation and dephasing for commutativity of $\Op H_S$ and $\Op H_{SB}$ 
we introduce a new mechanism, spin swap, to produce dephasing. This section also analyzes the effect of a finite bath energy spectrum in the
dissipation and dephasing processes. In sec. \ref{nv} the different dephasing mechanisms are shown in a more realistic scenario,
a diamond NV center. Finally, we present conclusions and future outlooks. 
\section{The Surrogate Hamiltonian}
\label{SH}
To study dephasing we use a complementary approach, the surrogate hamiltonian (SH) \cite{Roi}. The methods described previously begin with a reduced
description of the whole physical system. In contrast, the SH starts from a description of the total system and bath, yielding a model numerically feasible whose
validity is limited in time. The principal advantage of the method is that it goes beyond system-bath separability implied by the Markovian assumption.
The starting point is  Eq. (\ref{H}) and again the bath degrees of freedom are treated implicitly  by
abstract, representative modes. The core idea of the SH is truncation of the infinite modes of the bath by selecting representative modes, 
the modes that interacts intimately with the system \cite{Roi}. The truncation generates a new surrogate Hamiltonian generating the dynamics of the surrogate wave function. In the limit where SH 
includes an infinite number of modes coincides with the original Hamiltonian. To extend the convergence of the SH in the time domain, spin swaps can be applied to the bath, see Sec. \ref{s_swap}.
This swap operation leads to thermalization at long times and avoids spin reflections at the boundaries of the bath \cite{Giulia}. 
More details about the construction of the SH can be found in \cite{Roi, pump}. The truncation
leading to SH relies on the energy-time uncertainty principle. For a finite time $t\ll\infty$ the system only has time to interact with a finite number of bath modes $K\ll\infty$, 
performing unnecessarily a full density of states of the bath. 
Two important observations are derived from this argument. First, the SH is a suitable approach for ultra-short processes, 
and secondly, the number of required modes in the bath increases with the interaction strength. 
Intermediate and strong couplings require a computational effort. Nevertheless, no weak coupling assumption is required in the approach. 
\subsection{The bit representation}
The environment surrounding the primary system is considered to be a bath of TLS.
The Hilbert space $\mathcal{H}_B$ for the bath has dimension 
$2^K$ resulting from the combination of a single TLS with the rest of $K$ modes of the bath. 
To represent this space we choose the spin up $|1\rangle$ (TLS excited) and spin down $|0\rangle$ (TLS de-excited) base in the bit representation \cite{Roi, pump}
explained below.

The total Hilbert $\mathcal{H}_S\otimes\mathcal{H}_B$ space has the product dimension of $\mathcal{H}_S$ and $2^K$.
To represent a total state of $\Op H$ we need wave functions of dimension $2^KN_g$ if the system is described by a grid of dimension 
$N_g$. For example, if we consider a bath with $K=2$, two spins, the wave function spinor is
\beq
\Psi^{K=2}(q)=
\left(\begin{array}{c}
\psi_0(q,\phi)\\
\psi_1(q,\phi)\\
\psi_2(q,\phi)\\
\psi_3(q,\phi)
\end{array}\right)
\eeq
where $q$ represents the degrees of freedom of the system and $\phi$ the bath degrees of freedom. The spinor is bit ordered,
corresponding to a bit representation to each component of the spinor.
In the example considered, the bit representation of each spinor components is $0\rightarrow|00\rangle, 1\rightarrow|01\rangle, 2\rightarrow|10\rangle$ and
$3\rightarrow|11\rangle$, starting the count of bits from the right. The zeroth component corresponds to no bath modes excited, the first and second component to the excitation
of the first or second bath mode, and the third component to the simultaneous excitation of the first and second bath modes. 

The bath operators are sums over operators acting on a single mode. In the bit representation the operator of mode $k$ acts on the bit $k$ in the $2^K$ components
of the spinor. Note the difference between bath mode and spinor components. There are $K$ spin modes and $2^K$ spinor components. For example, the $k=0$ mode
that can be excited or de-excited should not be confused with the zeroth spinor component corresponding to all bath modes de-excited. 

It is useful to note that any bath operator can be written in terms of creation and annihilation operators. The creation operator of the $k$ mode is \cite{pump}
\beq
\Op\sigma_k^{{\dag}^{K}}=\prod_{i=1}^{K-k}\mathbb{I}_2\otimes\Op\sigma^{\dag}\otimes\prod_{i=1}^{k-1}\mathbb{I}_2,
\label{crea}
\eeq
where in the bit representation
\beq
\Op\sigma^{\dag}=
\left(\begin{array}{cc}
0&0\\
1&0
\end{array}\right).
\eeq
The annihilation operator for the $k$ mode is simply the conjugate expression of (\ref{crea}) built on $\Op\sigma$. The Pauli matrices for the 
$k$ modes are given by
\beq
\Op\sigma_k^x=\frac{\Op\sigma^{\dag}_k+\Op\sigma_k}{2}, \quad \Op\sigma_k^y=\frac{\Op\sigma^{\dag}_k-\Op\sigma_k}{2i}, \quad \Op\sigma_k^z=\frac{\Op\sigma^{\dag}_k\Op\sigma_k-\Op\sigma_k\Op\sigma^{\dag}_k}{2}.
\eeq
%
%
%
%
%
%
%
%
%
%
%
\begin{figure}[t]
\begin{center}
\includegraphics[width=0.3\linewidth]{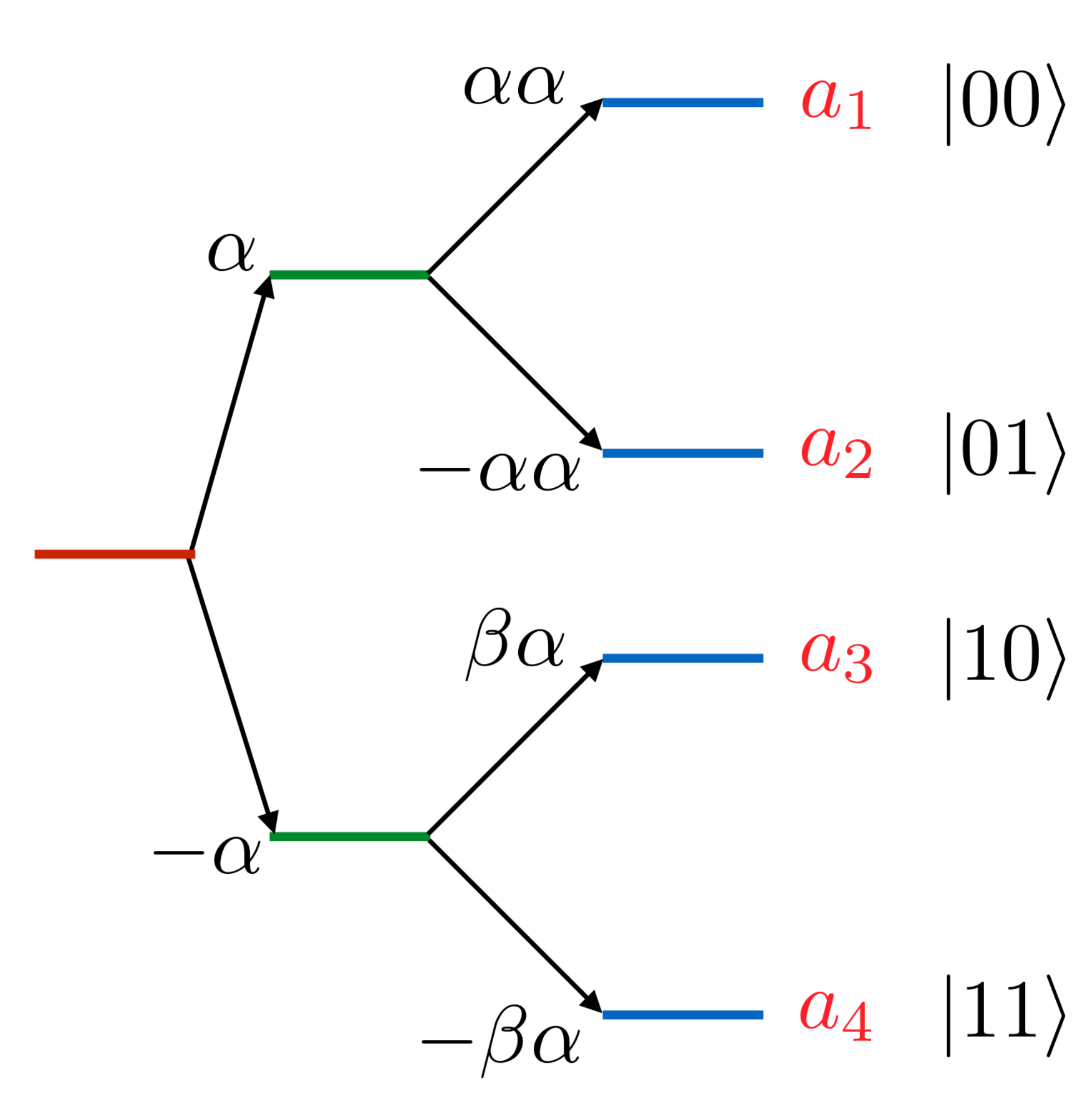}

\caption{Coding scheme of the spin bath. On the left: the branching code. On the right: the relative phases in the product base.}
\label{branching}
\end{center}
\end{figure}
%
%
%
%
%
%
%
%
%
%
\subsection{Spin swap}\label{s_swap}
An important operation in the stochastic surrogate hamiltonian is the swap of a bath spin by another \cite{gil}. The complication of the operation 
is to produce the swap keeping the entanglement between the changed spin with the remainder of the bath. The implementation is illustrated in the
following example. The wave function of two spins can be represented as
\beqa
|\phi\rangle&=&\lambda_1|00\rangle+\lambda_2|01\rangle+\lambda_3|10\rangle+\lambda_4|11\rangle  \nonumber\\
&=&\frac{1}{Z}(e^{a_1}|00\rangle+e^{a_2}|01\rangle+e^{a_3}|10\rangle+e^{a_4}|11\rangle),
\label{rep}
\eeqa
where $\lambda_k$ and $a_k\in\mathbb{C}$, and $Z^2=\sum_ke^{a_k+a_k^*}$ being $a^*_k$ the complex conjugate of $a_k$.
The irrelevant global phase is chosen in such a way $\sum_ka_k=0$. The relative phases can be related to the conditional amplitudes
using a branching tree \cite{gil}, see Fig. \ref{branching},
\beq
\left(\begin{array}{c}
a_1\\
a_2\\
a_3\\
a_4
\end{array}\right)
=
\left(\begin{array}{cccc}
1&1&0&0\\
1&-1&0&0\\
-1&0&1&0\\
-1&0&-1&0
\end{array}\right)
\left(\begin{array}{c}
\alpha\\
\alpha\alpha\\
\beta\alpha\\
0
\end{array}\right).
\label{A}
\eeq
The inverse relation reads
\beq
\left(\begin{array}{c}
\alpha\\
\alpha\alpha\\
\beta\alpha\\
0
\end{array}\right)
=
\left(\begin{array}{cccc}
0&0&-1/2&-1/2\\
1/2&-1/2&0&0\\
0&0&1/2&-1/2\\
1&1&1&1
\end{array}\right)
\left(\begin{array}{c}
a_1\\
a_2\\
a_3\\
a_4
\end{array}\right).
\label{Ai}
\eeq
To realize the swap operation we proceed as follows:
\begin{description}
\item[1.] Calculate the relative phases \{$a_k$\} from the computational coefficients \{$\lambda_k$\} using Eq. (\ref{rep}).
\item[2.] Using Eq. (\ref{Ai}) compute the coefficients $\alpha, \alpha\alpha$ and $\beta\alpha$ from \{$a_k$\}.
\item[3.] If the spin at left-side of the ket is swapped change $\alpha=b$ to the value $b$ of the swapped spin. If it
is the right-side spin then set $\alpha\alpha=\beta\alpha=b$.
\item[4.] Using Eq. (\ref{A}) recalculate the set \{$a_k$\} for the new branching coefficients $\alpha, \alpha\alpha, \beta\alpha$
\item[5.]  Calculate \{$\lambda_k$\} using again Eq. (\ref{rep}).
\end{description}
Usually in step 1 when we compute \{$a_k$\} from \{$\lambda_k$\} the phases do not satisfy $\sum_ka_k=0$, to solve it multiply and divide
$|\phi\rangle$ by $e^c$ with $c=-\frac{1}{2^K}\sum_k^K\log|\lambda_k|+i\arg(\lambda_k)$ and define $a_k=c+\log|\lambda_k|+i\arg(\lambda_k)$.
In step 3 we only modify the complex part of the $\alpha, \alpha\alpha, \beta\alpha$ to avoid wave function renormalization. 
 
The action of the swap operation is to produce a tensor product state between the swapped spin and the rest. For example, after swapping the right-side
spin of the ket we get
\beq
|\phi\rangle=(\lambda_1|0\rangle+\lambda_2|1\rangle)\otimes(\lambda_3|0\rangle+\lambda_4|1\rangle)
\eeq
with $\lambda_3/\lambda_4=e^{2b}$.

For a simulation containing spin swap operations we make $N_r$ realizations of the evolution to average the random swaps.
\section{Dephasing in the harmonic oscillator system}
\label{ho}
The generic example is the harmonic oscillator of mass $m$ and a constant frequency $\omega$ whose hamiltonian is given by
\beq
\Op H_s=\frac{\Op p^2}{2m}+\frac{1}{2}m\omega^2 \Op q^2
\label{sysoa}
\eeq
where $\Op p$ and $\Op q$ are the momentum and position quantum operators.

The bath contains $K$ two level systems with spins $1/2$ governed by the Hamiltonian
\beq
\Op H_B=\sum_k^K\epsilon_k\Op\sigma_k^{\dag}\Op\sigma_k,
\label{bathoa}
\eeq
where $\epsilon_k$ is the energy, and $\Op\sigma^{\dag}_k$ and $\Op\sigma_k$ are the creation and annihilation 
operators for the $k$ mode. The bath has a spectrum of modes with energies in the interval 
$\epsilon_k\in[\epsilon_0,\epsilon_c]$ with $\epsilon_c$ the cutting energy. For a bath
containing $K$ modes the interval is sampled selecting the discrete points $\epsilon_0 <\epsilon_1<...<\epsilon_{K-1}$.

The interaction between system and bath depends on the dissipative problem to be analyzed.
In addition to the classical decoherence, energy relaxation, quantum systems also show pure decoherence, phase relaxation. Energy relaxation is the exchange
of energy between the primary system and the bath which will eventually lead to thermal equilibrium. The hamiltonian
describing the energy transfer of the bath modes and the system has a dipolar nature and is given by
\beq
\Op H^r_{SB}=\Op q\otimes\sum_k^Kd_k(\Op \sigma^{\dag}_k+\Op \sigma_k)
\label{relax}
\eeq
where the coupling constants are given by $d_k=\sqrt{(J(\epsilon_k)/\rho(\epsilon_k))}$ with $J(\epsilon_k)$ the spectral
density and $\rho(\epsilon_k)$ the density of states. For all the examples in this paper we consider a linear spectral density
$J(\epsilon_k)=\eta\epsilon_k$ being $\eta$ a constant. The previous energy sampling  specifies a density of states for 
the discrete bath $\rho(\epsilon_k)\approx(\epsilon_{k+1}-\epsilon_k)^{-1}$.
The exchange energy process described by Eq. (\ref{relax})
can be understood as the subtraction of energy from the system and the creation of an excitation in a bath mode $\Op\sigma^{\dag}_k$
and the inverse process, a de-excitation of the bath mode $\Op\sigma_k$ that injects energy into the primary system.

The phase relaxation, or simply dephasing, occurs due to inelastic interactions between the primary system and different bath modes,
destroying the accumulated phase acquired by the system. 
%
%
%
%
%
%
%
%
%
%
%
%
%
%
%
\subsection{Standard pure dephasing}\label{pure_deph}
The most studied process for pure dephasing is due to the commutativity between $\Op H_S$ and $\Op H_{SB}$. It is governed by
\beq
\Op H^d_{SB}=\Op H_s\otimes\sum_{j<k}^Kc_{jk}(\Op \sigma_j^{\dag}\Op \sigma_k+\Op \sigma_k^{\dag}\Op \sigma_j),
\label{dep}
\eeq 
where the coupling constants $c_{jk}$ are 
\beq
c_{jk}=\frac{c}{K(K-1)}e^{\frac{(\epsilon_j-\epsilon_k)^2}{2\sigma^2_{\epsilon}}},
\eeq
being $c$ the global dephasing parameter and $\sigma_{\epsilon}$ the inelastic bias. Due to the commutation $[\Op H^d_{SB},\Op H_B]=0$ the number of bath excitations is not changed and dephasing occurs 
if the bath is initially {\it activated} i.e., initially excited.
Note also that each of the parts $\Op H_S, \Op H_B$, and $\Op H_{SB}$ conmute with the total hamiltonian $\Op H$, system, bath and interaction energies remain 
constant during the process. A physical interpretation of Eq. (\ref{dep}) is the excitation of a bath mode at the expense of the de-excitation of another mode
and vice versa. To modulate the excitation of the primary system the bath modes have to be almost degenerate and their frequencies on-resonance
with the system frequency. To model pure dephasing processes in this system let us consider the total hamiltonian
%
%
%
%
%
\begin{figure}[t]
\begin{center}
\includegraphics[width=0.49\linewidth]{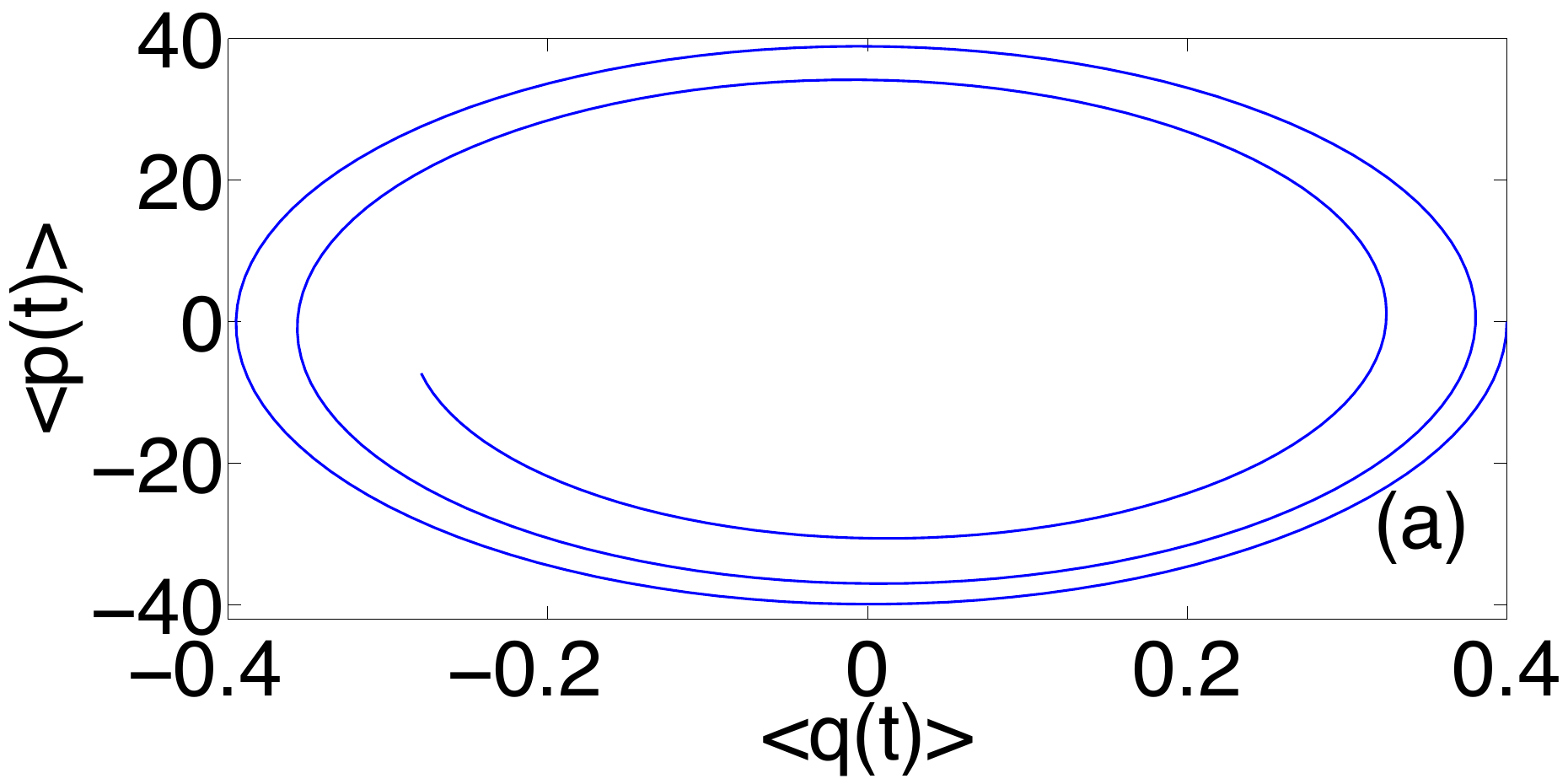}
\includegraphics[width=0.49\linewidth]{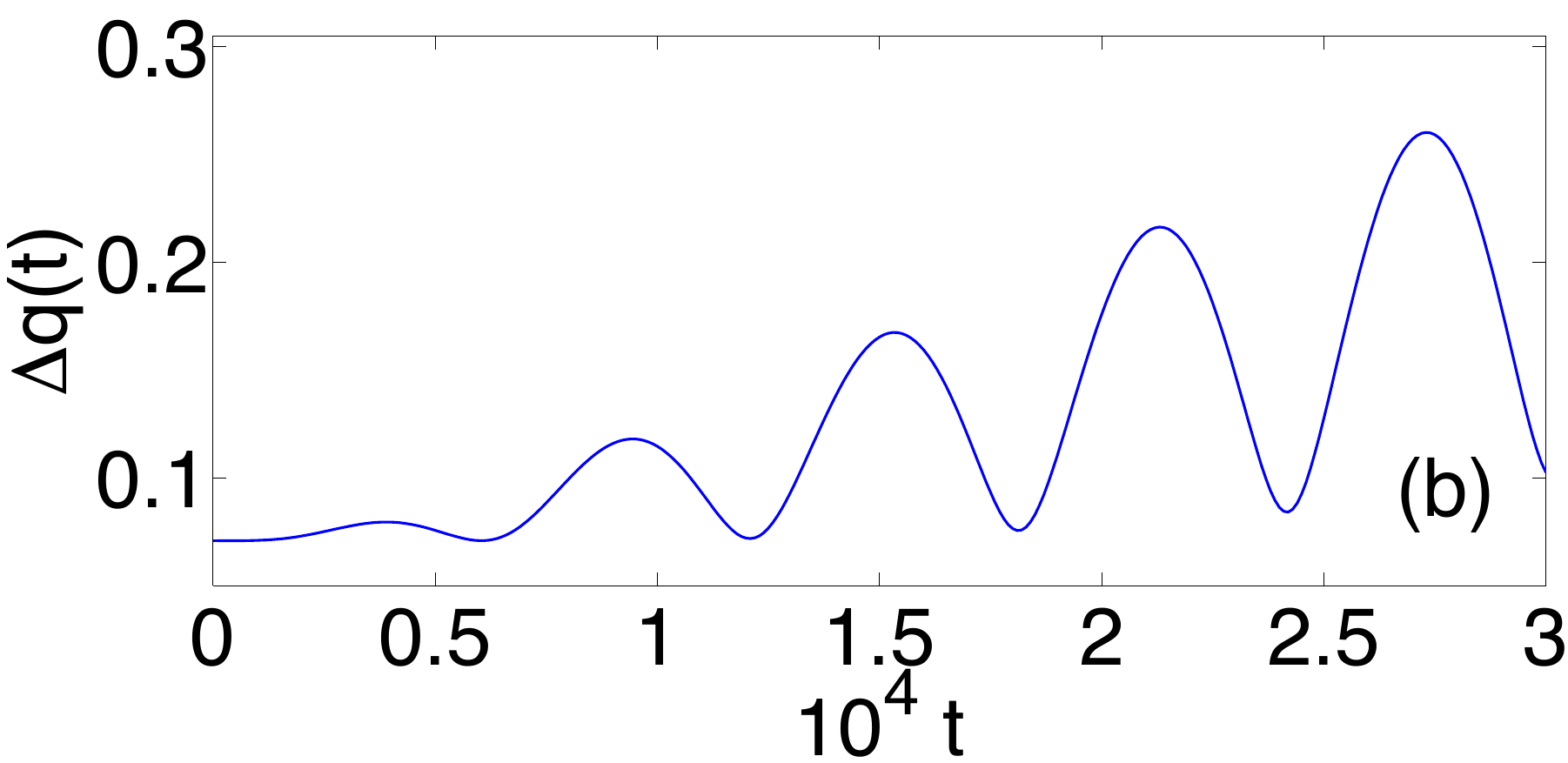}
\includegraphics[width=0.49\linewidth]{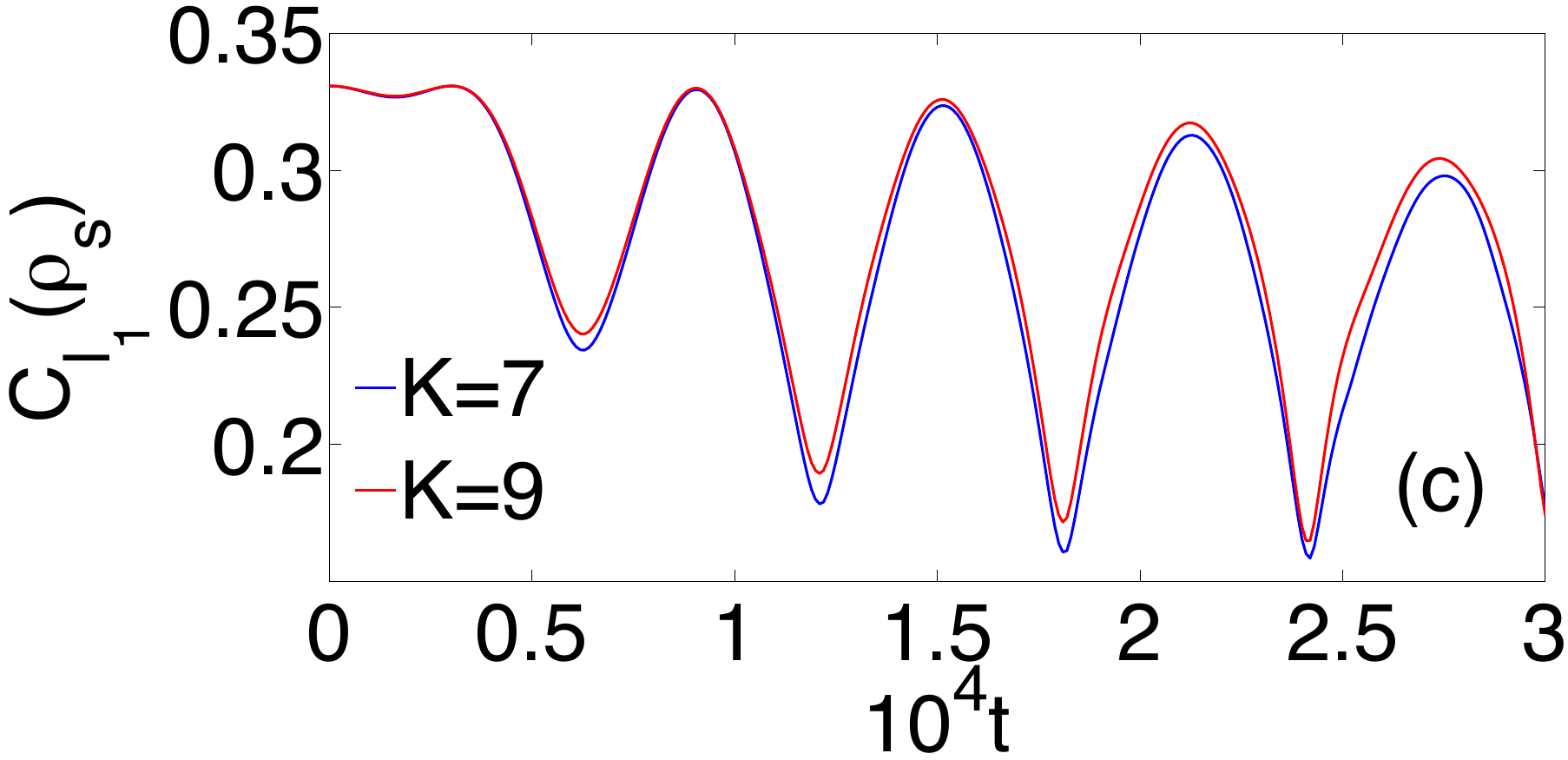}
\includegraphics[width=0.49\linewidth]{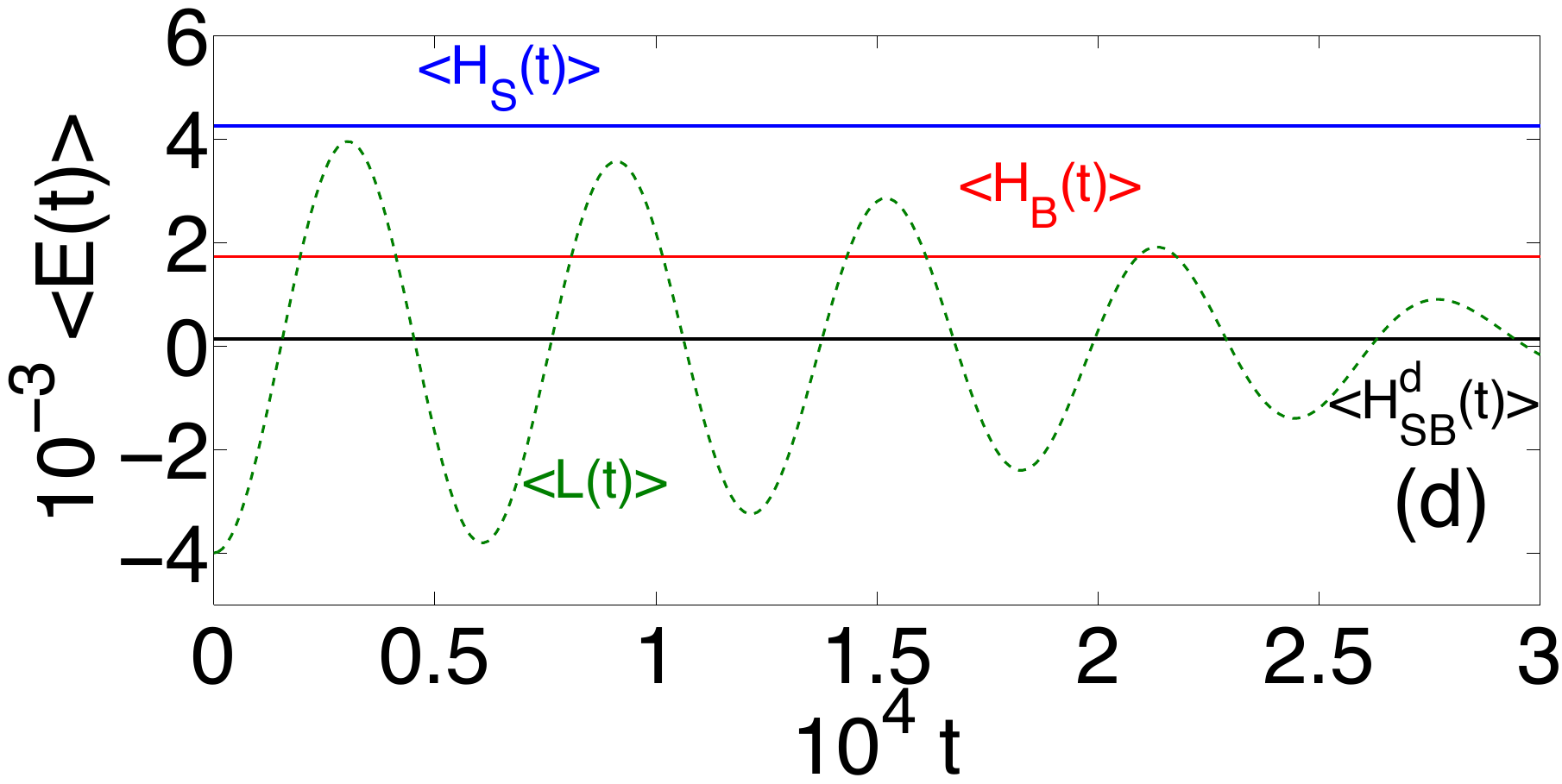}

\caption{
(a) Wave packet evolution in phase space under the dynamics generated by Eq. (\ref{pd}). At initial time it is placed at (0.4,0) and starts to rotate around
(0,0). The wave packet follows a spiral instead of a closed circle because of pure dephasing.
(b) Standard deviation of position $\Delta \Op q(t)=\sqrt{\langle \Op q^2(t)\rangle-\langle \Op q(t)\rangle^2}$ as a function of time. During the evolution the wave
packet spreads increasing the uncertainty in the position and momentum. 
(c) Coherence
function ${\mathcal C}_{l_{1}}(\Op\rho_S)=\int\int dqdq'|\langle q|\Op\rho_S(t)|q'\rangle|$ for $q\neq q'$ as a function of time. Coherence spreads and decays during the evolution.
(d) Energy mean values as a function of time. 
The relations $[\Op H,\Op H_S]=[\Op H,\Op H_B]=[\Op H,\Op H_{SB}^d]=0$ indicate  that the system and bath energies are constant in the pure dephasing process. However, the Lagrangian
mean value of the primary system $\langle \Op L(t)\rangle=\langle\Psi(t)|\Op p^2/(2m)-m\omega^2\Op q^2/2|\Psi(t)\rangle$
does not remain constant indicating that $|\Psi(t)\rangle$ is a non-stationary state of $\Op H_S$. 
Parameter values: $K=9$, $m=2\cdot 10^{5}$, $\omega=5\cdot 10^{-4}$, $c=0.5$, $\sigma_{\epsilon}=5\cdot 10^{-6}$, 
$\Delta\omega=0.01\omega$, and $d=0.4$
}
\label{DEP}
\end{center}
\end{figure}
%
%
%
%
%

%
%
%
%
%
%
\beq
\label{pd}
\Op H=\frac{\Op p^2}{2m}+\frac{1}{2}m\omega^2\Op q^2+\sum_k^K\epsilon_k\Op \sigma_k^{\dag}\Op \sigma_k+\bigg(\frac{\Op p^2}{2m}+\frac{1}{2}m\omega^2\Op q^2 \bigg) \otimes \sum_{j<k}^Kc_{jk}(\Op \sigma_j^{\dag}\Op \sigma_k+\Op \sigma_k^{\dag}\Op \sigma_j), 
\eeq
%
%
%
%
%
%
%
%
%
%
%
%
%
\begin{figure}[t]
\begin{center}
\includegraphics[width=0.32\linewidth]{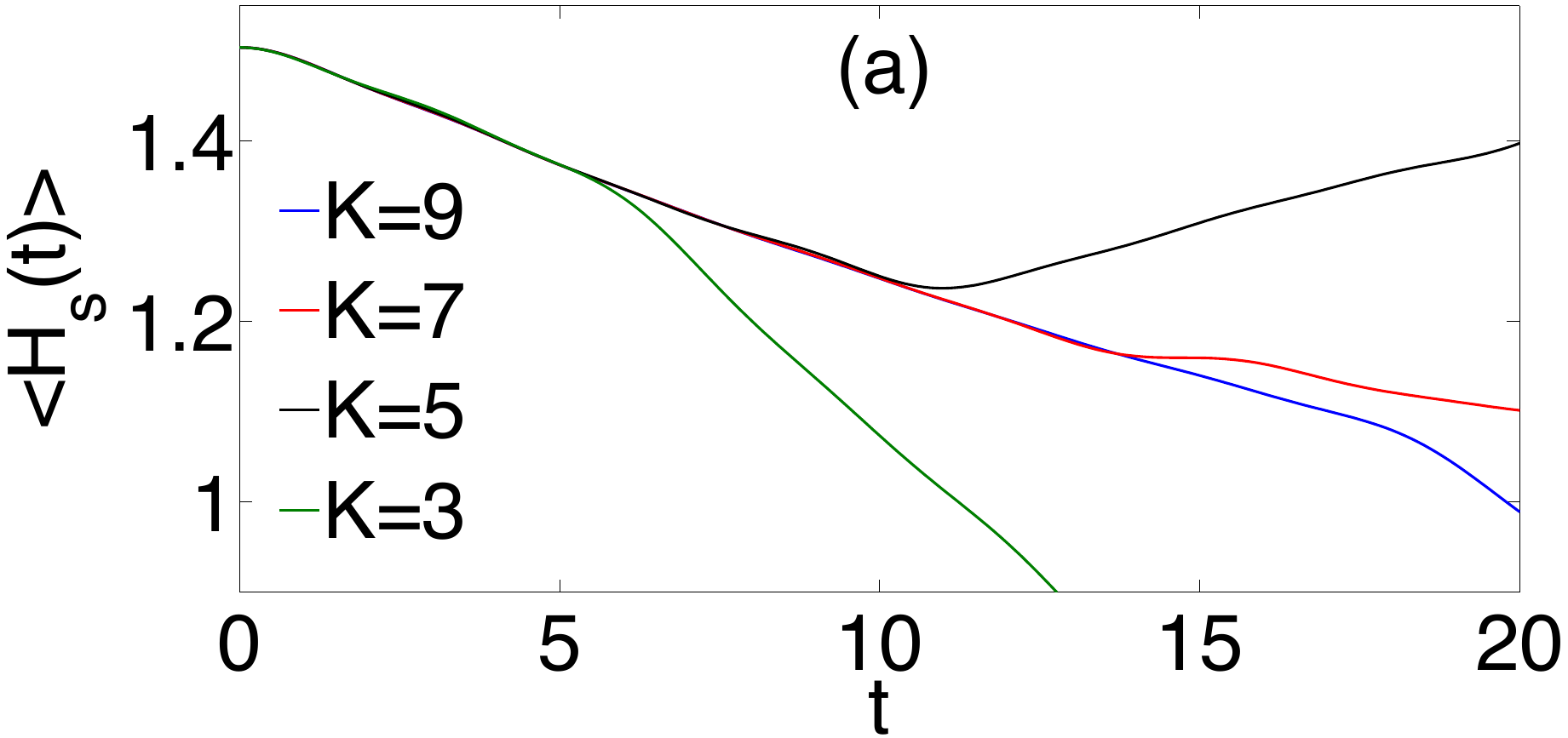}
\includegraphics[width=0.32\linewidth]{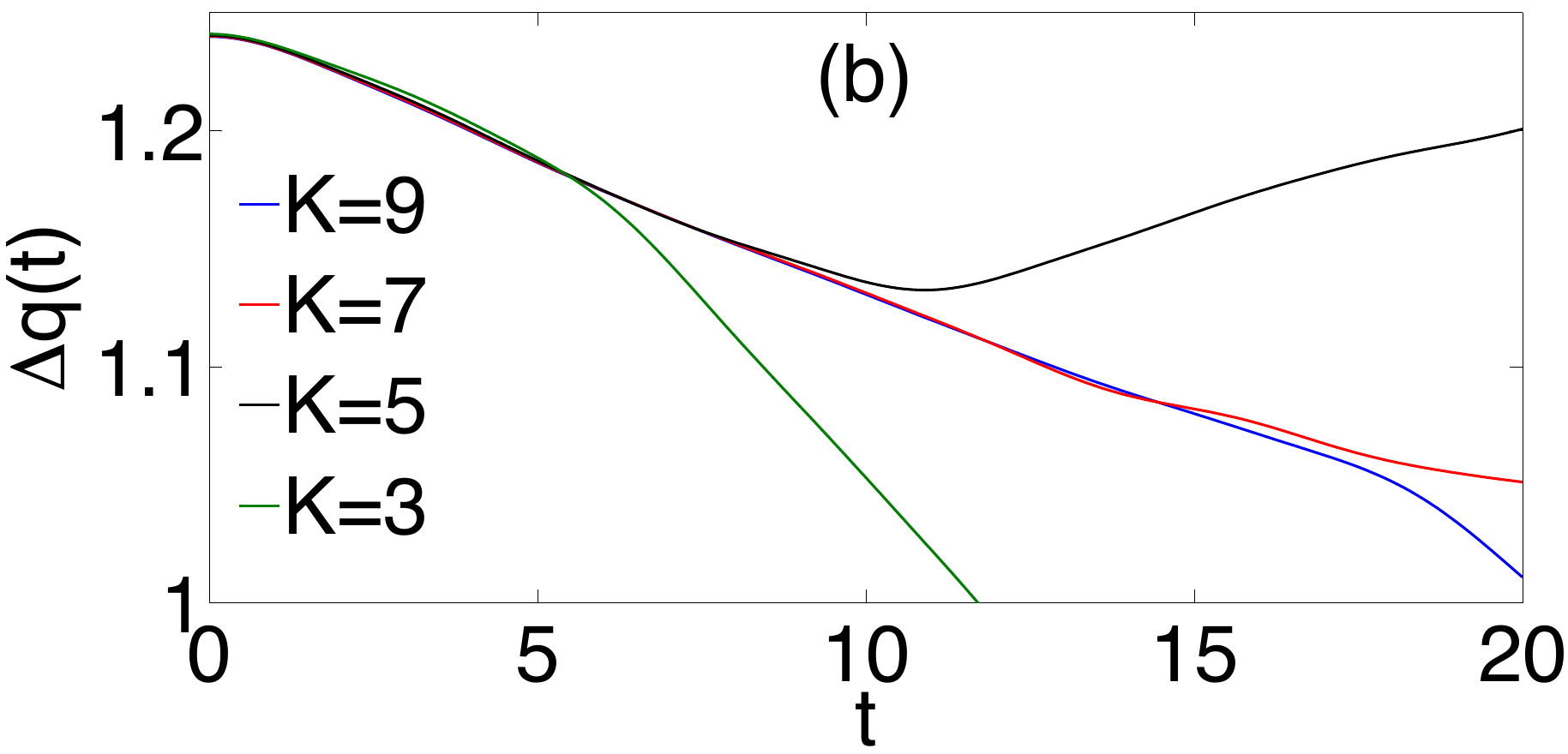}
\includegraphics[width=0.32\linewidth]{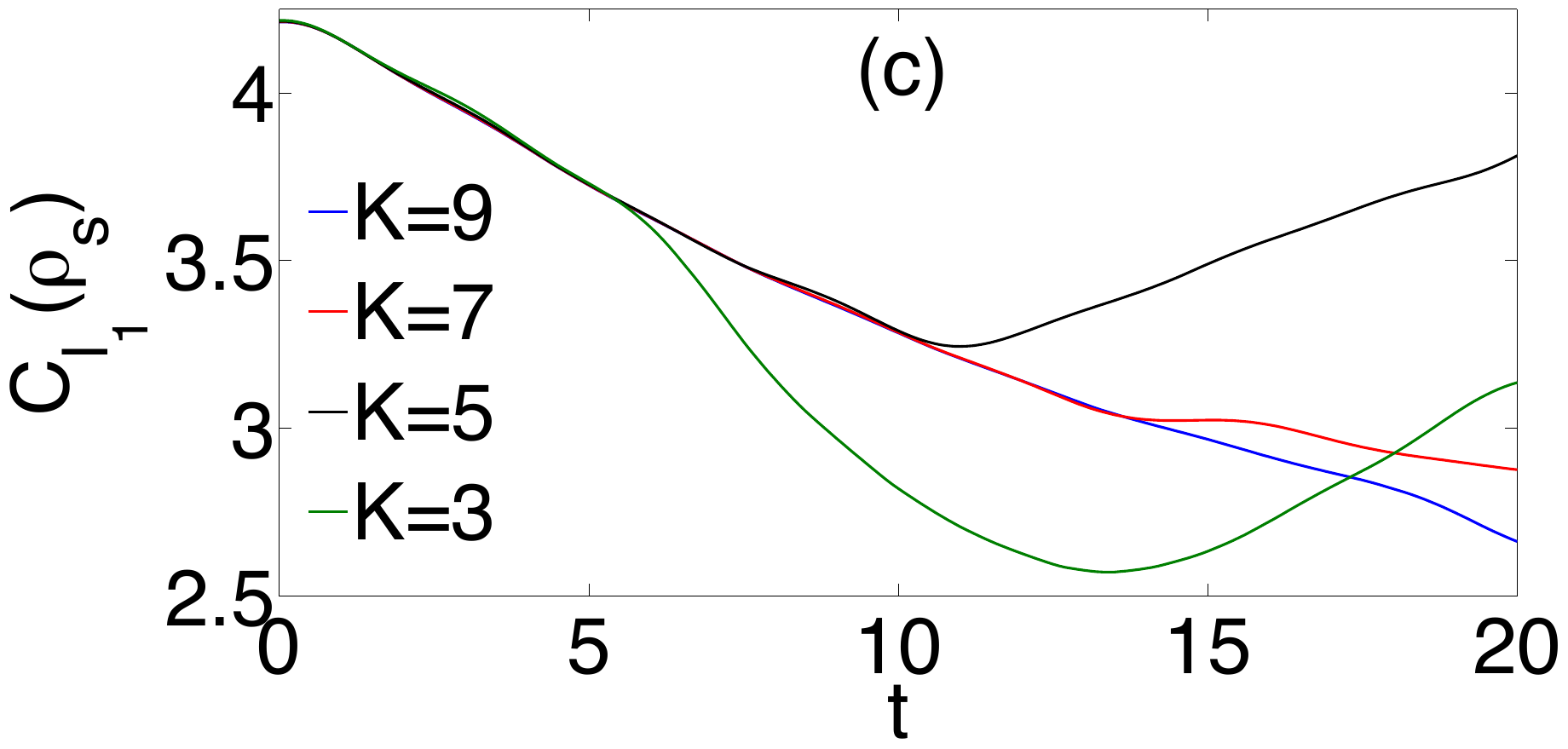}

\caption{(a) Energy relaxation for an infrared excitation of (\ref{Hdis}). The ground state $|\Psi_g\rangle$ of the total hamiltonian
was computed with imaginary time evolution. Increasing the number of bath modes the convergence of results is prolonged. (b) 
Standard deviation of the position as a function of time for different bath modes showing also an exponential decay. (c) Coherence
function ${\mathcal C}_{l_{1}}(\Op\rho_S)$ as a function of time showing an exponential decay
of coherence. Parameter values: $m=\omega=1$, $\epsilon_0=0$, $\epsilon_c=3$, $\eta=10^{-2}$.
}
\label{dis}
\end{center}
\end{figure}
where the energy spectrum of the bath modes is in the interval $\epsilon_k\in[\omega-\Delta\omega,\omega+\Delta\omega]$ ($\Delta\omega\ll\omega$) to be on-resonance with the harmonic oscillator.
The initial state $|\Psi(0)\rangle$ was chosen such that the system is in its ground state and the bath is randomly excited. This state $|\Psi(0)\rangle$ was displaced 0.4 a.u. and evolved in time. 
The state $|\Psi(t)\rangle$ rotates around $\langle \Op q(t)\rangle=\langle\Psi(t)|\Op q|\Psi(t)\rangle=0$, $\langle \Op p(t)\rangle=0$ point in the phase space. Due to pure dephasing the wave packet also spreads
during the revolution thus the wave packet does not complete a closed circle but a spiral one, see Fig. \ref{DEP} (a). Consequently the uncertainty of $\Delta\Op q(t)$ and $\Delta\Op p(t)$ increase in time
destroying the coherence of the primary system. Figures \ref{DEP} (b) and (c) show how the initial state, displaced ground state of the harmonic oscillator $\Delta\Op q(0)\approx\sqrt{1/(2m\omega)}$, spatially spreads
due to pure dephasing. The oscillation period of $\Delta\Op q$ and ${\mathcal C}_{l_{1}}(\Op\rho_S)$ corresponds to the revolution of the wave packet in phase space, for example the maximums 
of the coherence function are $\langle \Op q(t)\rangle=0$ in Fig. \ref{DEP} (a) whereas the minimums $\langle \Op p(t)\rangle=0$.
As we stated before, the commutativity $[\Op H_{SB}^d,\Op H_S]=[\Op H_{SB}^d,\Op H_B]=0$ assures that pure dephasing does not change the energy of the system 
and bath, see Fig. \ref{DEP} (d). This fact does not mean that the state $|\Psi(t)\rangle$ is an eigenvector of $\Op H_S$, for example if we compute the Lagrangian mean value $\langle \Op L(t)\rangle$
it is not constant indicating that the wave packet is not stationary.

\subsection{Spin swap dephasing}\label{swap_deph}
To study a second mechanism to produce pure dephasing, we consider the standard model used to analyze dissipative dynamics of the
harmonic oscillator linearly coupled to a spin bath, the hamiltonian is
\beq
\Op H=\frac{\Op p^2}{2m}+\frac{1}{2}m\omega^2 \Op q^2+\sum_k^K\epsilon_k\Op \sigma_k^{\dag}\Op \sigma_k+\Op q\otimes\sum_k^Kd_k(\Op \sigma^{\dag}_k+\Op \sigma_k),
\label{Hdis}
\eeq
where the bath has an energy spectrum $\epsilon_k\in[0,3\omega]$ and initially is de-excited. 
For an initial state $|\Psi(0)\rangle=\Op q|\Psi_g\rangle$, corresponding to an infrared excitation of the ground state, it is well known that the energy system
$\langle \Op H_S(t)\rangle$ decays exponentially with a rate $2\pi\eta\omega$ predicted analytically for weak coupling $\eta$ \cite{Louisell}, see Fig. \ref{dis}.  
The coherence ${\mathcal C}_{l_{1}}(\Op\rho_S)$ and the spatial standard deviation 
$\Delta \Op q(t)$ also show an exponential decay when the system decays from the excited $\Delta \Op q(0)\approx\sqrt{ 3/(2m\omega)}$ to the ground state $\Delta \Op q\approx\sqrt{ 1/(2m\omega)}$
of the harmonic oscillator.
%
%
%
%
%
\begin{figure}[t]
\begin{center}
\includegraphics[width=0.32\linewidth]{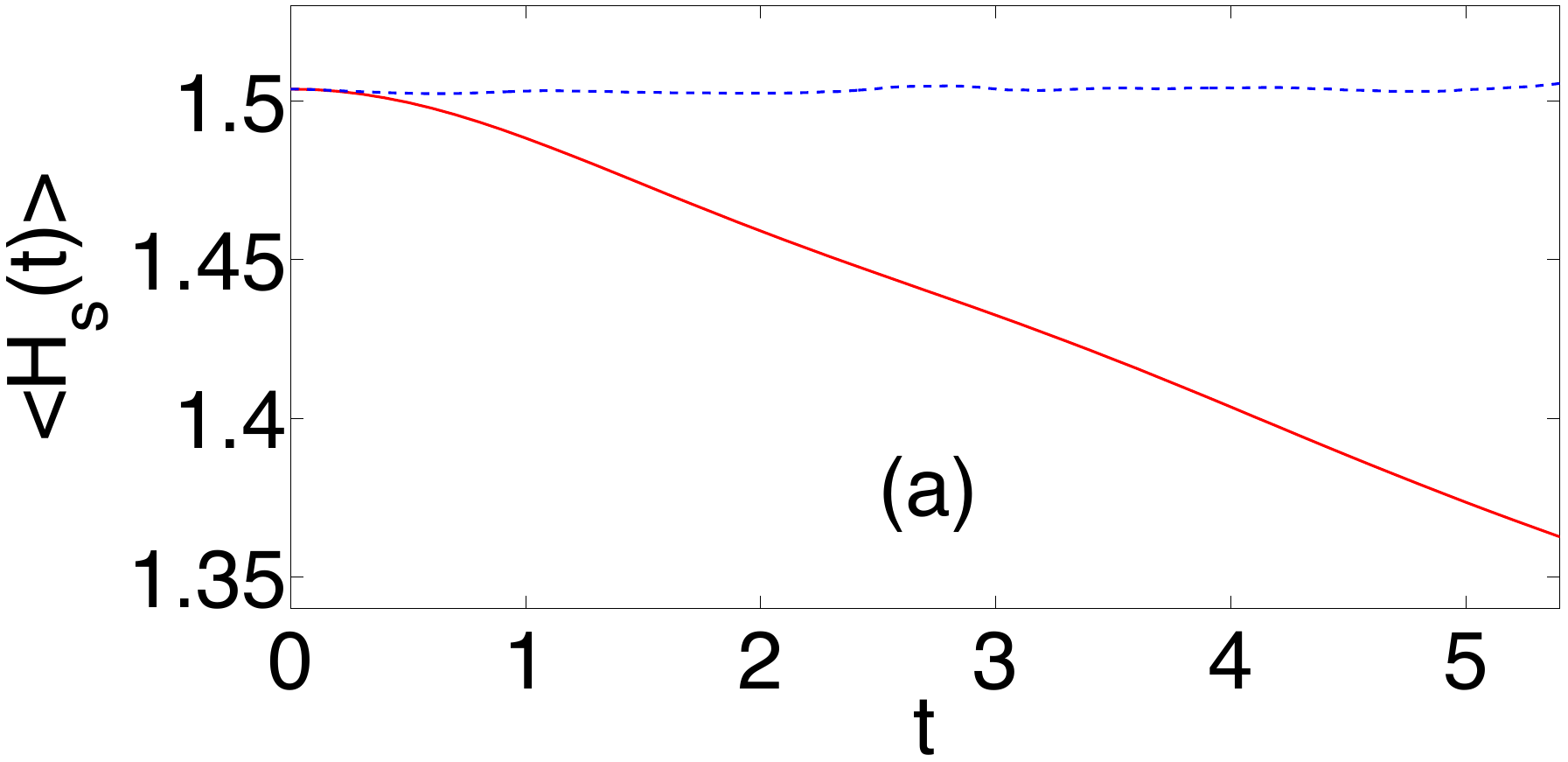}
\includegraphics[width=0.32\linewidth]{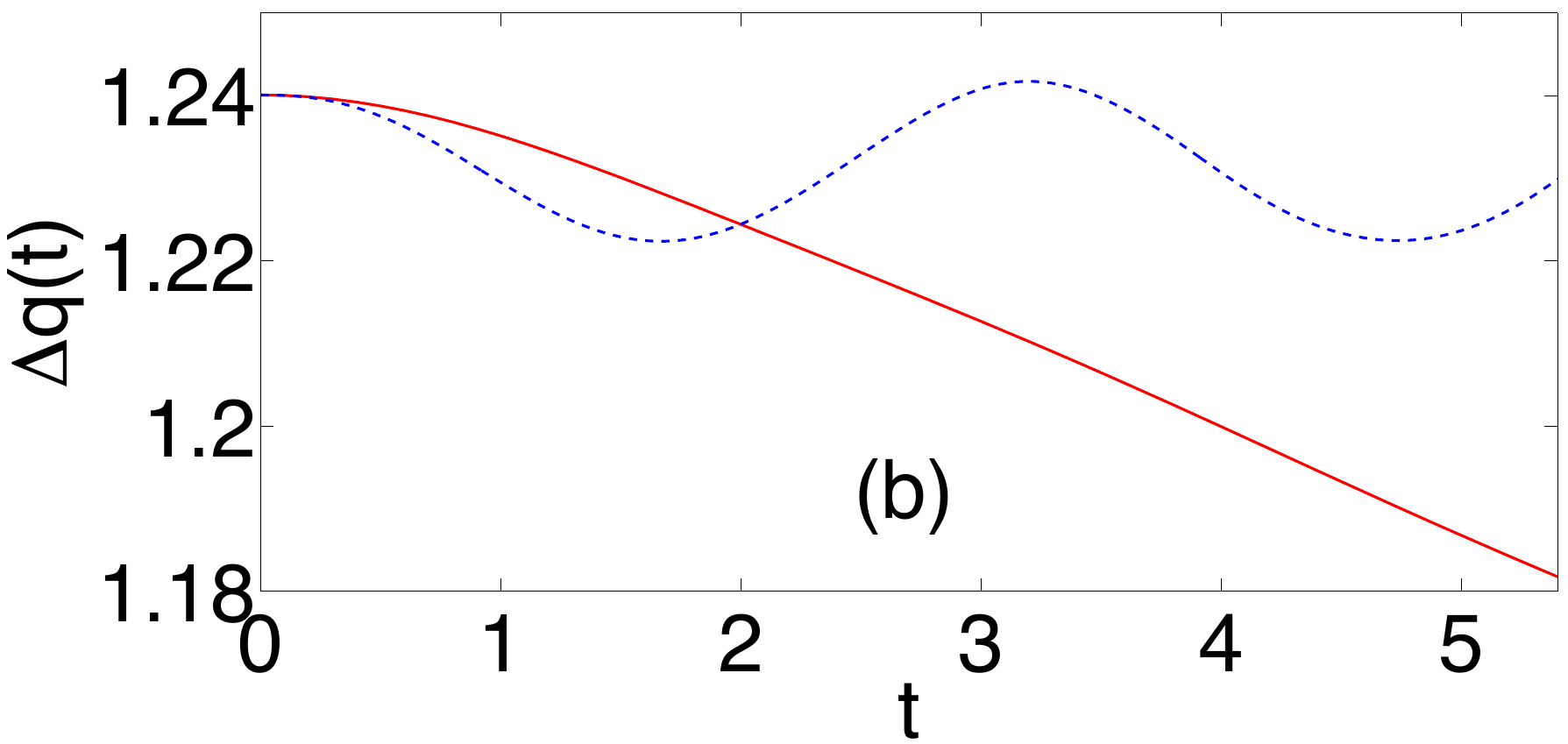}
\includegraphics[width=0.32\linewidth]{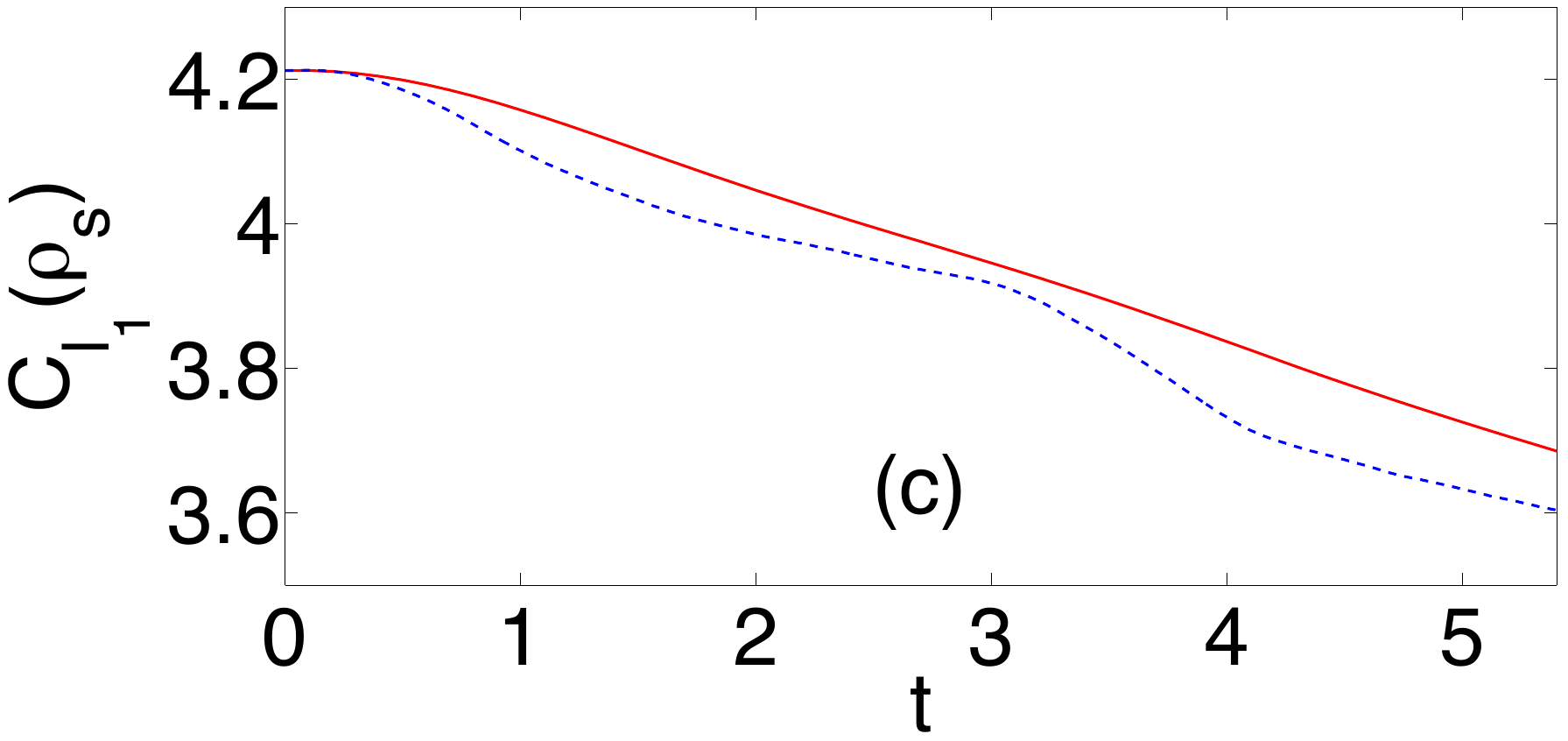}

\caption{(a) System energy evolution as a function of time for an infrared excitation of (\ref{Hdis}). When spin swap is produced faster than $t_Z$ the system does not
decay and its energy remains constant (blue-dashed line). If swap is absent the system decays exponentially fast (red-solid line, same as in Fig. \ref{dis}). (b) Standard
deviation of the position as a function of time. Applying spin swap the wave packet evolves spreading in time around $\Delta\Op q\approx\sqrt{3/(2m\omega)}$ corresponding to the
system excited state (blue-dashed line). If swap is absent the standard deviation decays exponentially (red-solid line, same as in Fig. \ref{dis}). (c) Coherence function $\mathcal{C}_{l_{1}}(\Op\rho_S)$
as a function of time. Parameters: $K=7$, $N_r=50$, and the rest are the same as in Fig. \ref{dis}.
}
\label{ho_swap}
\end{center}
\end{figure}
%
%
%
%
%
This model can exhibit pure dephasing if the bath experiments spin swap with a rate $t_f$ faster than the Zeno time $t_Z$ of the system. 
The evolution of the initial state $|\Psi(0)\rangle$ under (\ref{Hdis}) is governed by the evolution operator $\Op U(t)=exp(-i\Op Ht)$. The ``survival probability'' at time
$t$ is $P(t)=|\langle\Psi(0)|\Psi(t)\rangle|^2=|\langle\Psi(0)|\Op U(t)|\Psi(0)\rangle|^2=1-t^2/t_Z^2$ with
\beq
t_Z=1/\Delta \Op H=\bigg(\langle\Psi(0)| \Op H^2|\Psi(0)\rangle-\langle\Psi(0)| \Op H|\Psi(0)\rangle^2\bigg)^{-1/2}.
\eeq
The quantum Zeno effect \cite{Zeno} is the suppression of unitary time evolution by quantum 
decoherences produced not only by continuous measurements of the evolving state $|\Psi(t)\rangle$ but also by interactions of the primary system with the bath, 
stochastic fields, etc. If we look at the effect from the frame that moves in such a way the unitary evolution is cancelled,
the system, otherwise stationary, is guided by the decoherence processes. In our model, the initial infrared excitation of (\ref{Hdis}) is frozen due to the system-bath interactions performing
spin swaps in the bath. The swaps suppress the exponential decay of the system keeping the system energy constant as in the pure dephasing
process. The frozen state is not simply a stationary state of Eq. (\ref{Hdis}) as we infer from Fig. \ref{ho_swap}, its spatial standard deviation spreads in time around 
$\Delta \Op q\approx\sqrt{ 3/(2m\omega)}$, corresponding to the excited state.
\subsection{Off-resonance dephasing}
\begin{figure}[t]
\begin{center}
\includegraphics[width=0.49\linewidth]{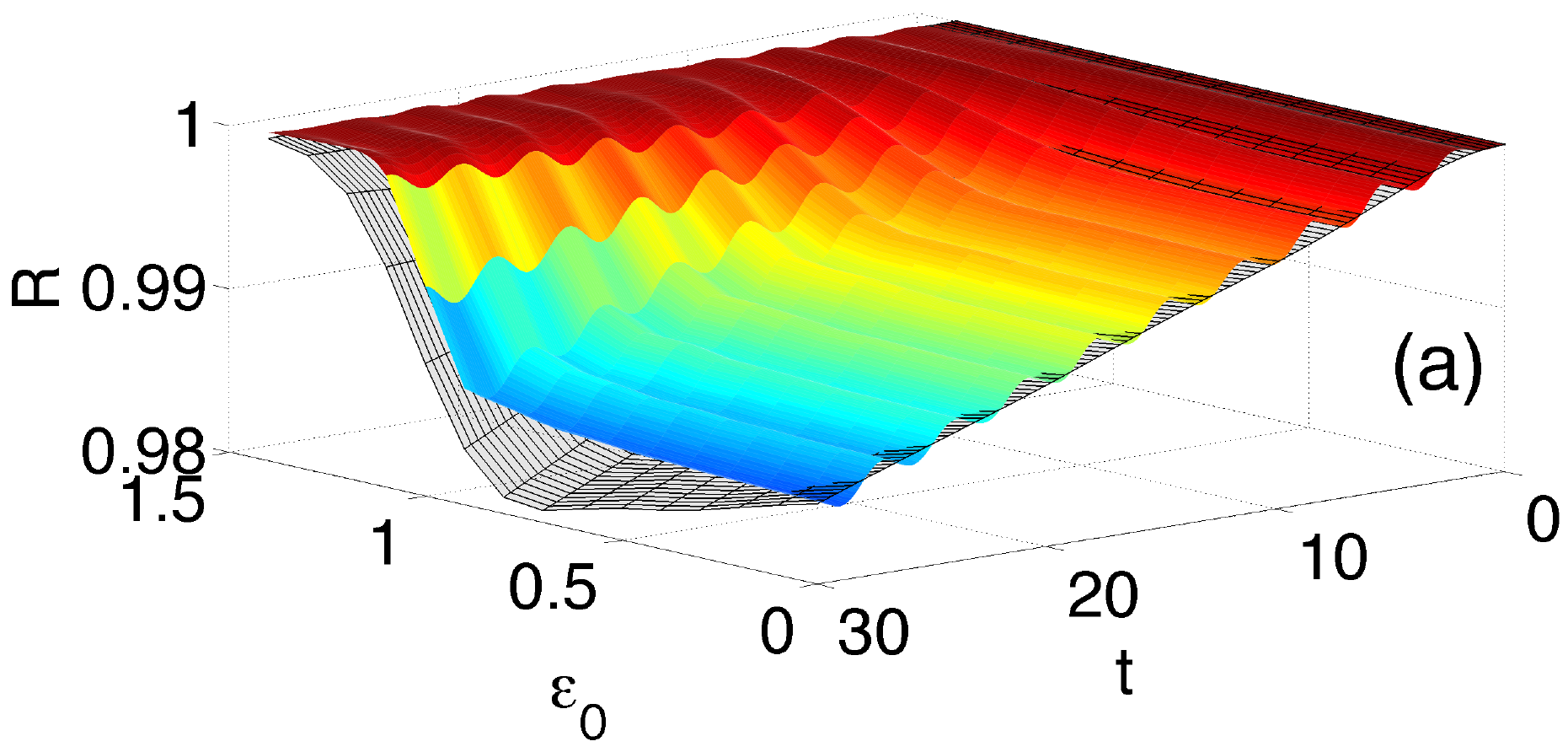}
\includegraphics[width=0.49\linewidth]{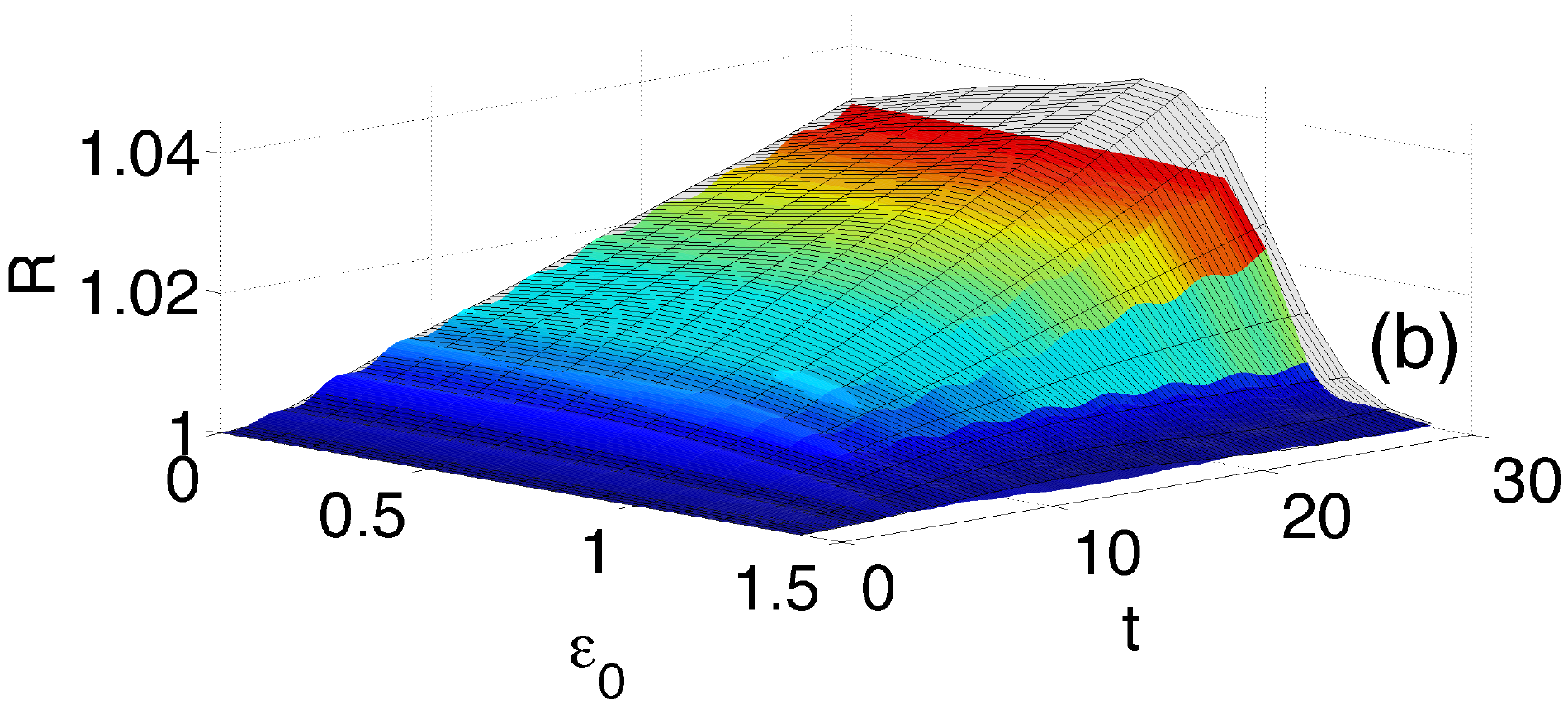}

\caption{Ratio between energy relaxation and dephasing as a function of time for different bath spectra configurations. (a) The initial state is a
displaced infrared excitation of the total hamiltonian ground state  $|\Psi(0)\rangle=e^{-i\hat pd}\Op q|\Psi_g\rangle$. (b) The initial state
is the displaced ground state of the total hamiltonian  $|\Psi(0)\rangle=e^{-i\Op pd}|\Psi_g\rangle$. The  black surface coresponds to the solution
for the harmonic oscillator bath computed with Eq. (\ref{ra_ap}).
Parameter values: $K=11$, $m=\omega=1$, $\epsilon_c=1.5$, $\eta=10^{-3}$, and $d=0.4$. (a) $\Op H(0)=1.58$, (b) $\Op H(0)=0.58$. 
}
\label{ratios}
\end{center}
\end{figure}

As pointed out in the introduction a possible mechanism for pure dephasing is the separation of time scales between system and bath produced by an energy gap.
In this section we analyze if the dipolar coupling (\ref{relax}) can produce pure dephasing
through this mechanism and the effect of a finite bath frequency spectrum in the energy relaxation and dephasing processes. To this end we employ again the same hamiltonian as
in the previous sections
\beq
\Op H=\frac{\Op p^2}{2m}+\frac{1}{2}m\omega^2\Op q^2+\sum_k^K\epsilon_k\Op \sigma_k^{\dag}\Op \sigma_k+\Op q\otimes\sum_k^Kd_k(\Op \sigma^{\dag}_k+\Op \sigma_k),
\label{off}
\eeq
where now different energy spectra of the bath modes $\epsilon_k\in[\epsilon_0,\epsilon_c]$ are considered. 
We kept constant the cutting frequency $\epsilon_c=1.5\omega$ and $\epsilon_0$ takes values from $0$ to $1.4\omega$. In contrast with the QME and 
Lindblad formalisms that assume an infinite bath spectra and consequently always there is a frequency of the bath on resonance with the system,
we can now study off-resonance processes and the finite effect of a bath spectrum. It is well known that in energy relaxation processes
also dephasing occurs, however the inverse statement is not always true as we saw in Secs. \ref{pure_deph} and \ref{swap_deph}, where in the pure dephasing processes system and bath 
do not exchange energy. To analyze the effect of the bath finite spectrum in the energy relaxation and dephasing processes we define the ratio
\beq
R(t)=\frac{\langle\Op a(0)\rangle}{\langle\Op H_S(0)\rangle}\frac{\langle\Op H_S(t)\rangle}{\langle\Op a(t)\rangle},
\eeq
for $R(t)<1$ processes the energy exchange rate is faster than dephasing and vice versa. The results are presented in Fig. \ref{ratios}. 
%
%
%
%
%
%
%
%
%
%
%
%
%
%

To gain analytical insight and understand this figure let us modify our initial model constituted by a harmonic oscillator coupled
to a spin bath and cast the environment by a bath constituted by harmonic oscillators instead. First we define $\epsilon_k:=\omega_k$,  $\chi_k:= d_k/\sqrt{2m\omega_k}$,
 $\Op a=\sqrt{m\omega/2}[\Op q+i\Op p/(m\omega)]$,  $\Op a^{\dag}=\sqrt{m\omega/2}[\Op q-i\Op p/(m\omega)]$,
 $\Op\sigma_k^{\dag}:=\Op b_k^{\dag}$, and $\Op\sigma_k:=\Op b_k$ where $\Op a, \Op a^{\dag}, \Op b_j, \Op b_j^{\dag}$, are the boson annihilation and creation operators of the oscillator and the $j$th bath mode.
 Replacing into Eq. (\ref{off}) after assuming the rotating wave approximation $(\Op a^{\dag}\Op\sigma_k^{\dag}+
\Op a\Op\sigma_k\ll\Op a^{\dag}\Op\sigma_k+\Op a\Op\sigma_k^{\dag})$ we get
\beq
\Op H=\omega\bigg(\Op a^{\dag}\Op a+\frac{1}{2}\bigg)+\sum_{k=1}^K\omega_k\Op b_k^{\dag}\Op b_k+\sum_{k=1}^K\chi_k(\Op b_k\Op a^{\dag}+\Op b_k^{\dag}\Op a).
\label{Hho}
\eeq
The advantage of this new hamiltonian is that it can be completely solved analytically \cite{braun}. For convenience, we shift the zero energy point to cancel the $\omega/2$ term 
and define $\Op c_0=\Op a$ and $\Op c_k=\Op b_k$ to rewrite 
\beq
\label{cHc}
\Op H=\mathds{C}^{\dag}\mathds{H}\mathds{C}
\eeq
where
\beq
\mathds{H}=
\left(\begin{array}{ccccc}
\omega&\chi_1&\chi_2&\cdots&\chi_K\\
\chi_1&\omega_1&0&\cdots&0\\
\chi_2&0&\omega_2&\cdots&0\\
\vdots&\vdots&\vdots&\ddots&\vdots\\
\chi_K&0&0&\cdots&\omega_K
\end{array}\right), 
\quad \mathds{C}=
\left(\begin{array}{c}
\Op c_0\\
\Op c_1\\
\vdots\\
\Op c_K
\end{array}\right).
\label{Hmatrix}
\eeq
Let us assume that $\mathds{A}$ is the unitary transformation that diagonalizes $\mathds{H}$,  
\beq
\lambda=\mathds{A}^{\dag}\mathds{H}\mathds{A}=
\left(\begin{array}{cccc}
\lambda_0&0&\cdots&0\\
0&\lambda_1&\cdots&0\\
\vdots&\vdots&\ddots&\vdots\\
0&0&\cdots&\lambda_K
\end{array}\right).
\label{eigenval}
\eeq
The column vector $\mathds{C}$ transforms as $\mathds{B}=\mathds{A}^{\dag}\mathds{C}$ and then the hamiltonian Eq. (\ref{cHc}) reads, $\Op H=\mathds{B}^{\dag}\lambda\mathds{B}$.
In this representation, the Heisenberg equations of motion for the $\mathds{B}$ operators are 
\beq
i\frac{d\mathds{B}}{dt}=[\mathds{B},\Op H]=\lambda\mathds{B},
\eeq
whose solution is 
\beq
\mathds{B}(t)=e^{-i\lambda t}\mathds{B}(0),
\eeq
where the exponential of the diagonal matrix $\lambda$ is defined by its series expansion. Inverting the transformation $\mathds{A}$ we find the evolution of the original operators,
\beq
\mathds{C}(t)=\mathds{U}(t)\mathds{C}(0),
\eeq
 where the evolution operator of our system  is given by $\mathds{U}(t)=\mathds{A}e^{-i\lambda t}\mathds{A}^{\dag}$. In particular for the boson annihilation operator of the system
 we have
 \beq
 \label{a}
 \Op a(t)=U_{00}(t)\Op a(0)+\Op G(t)
 \eeq
 where 
 \beq
 \label{G}
 \Op G(t)=\sum_{j=1}^K U_{0j}(t)\Op b_j(0)
 \eeq
 is a function expressed only in terms of the initial state of the bath. Taking the complex conjugate of expressions (\ref{a}) and (\ref{G}) we get a similar expression for the creation operator of the system.
 The system energy $\Op H_S(t)=[\Op a^{\dag}(t)\Op a (t)+1/2]\omega$ can be expressed as
 \beq
 \label{Es}
 \Op H_S(t)=|U_{00}(t)|^2\bigg(\Op H_S(0)-\frac{\omega}{2}\bigg)+\omega\bigg[U_{00}^*(t)\Op G(t)\Op a^{\dag}(0)+U_{00}(t)\Op G^{\dag}(t)\Op a(0)+\frac{1}{2}\bigg].
 \eeq
Once we have Eqs. (\ref{a}) and (\ref{Es}) an exact analytical expression for $R(t)$ can be obtained, however due to the different essence of the spin and harmonic baths this exact expression
does not reproduce exactly the oscillations of Fig. \ref{ratios}. However, we can get an expression that fits accurately the envelope of the figure setting $\Op G(t)=\Op G(t)^{\dag}=0$,
\beq
\label{ra_ap}
R(t)=|U_{00}(t)|-\frac{\omega}{2\Op H_S(0)}\bigg( |U_{00}(t)|-\frac{1}{|U_{00}(t)|}\bigg),
\eeq
where the initial condition $\Op H_{S}(0)$ is deduced from the spin bath model. For an explicit expression of $|U_{00}(t)|$ in terms of the eigenvalues and eigenvectors see {\ref{U00}}. It is a positive
function that at $t=0$ takes the maximum value $|U_{00}(0)|=1$ as a function of time. 
Figure \ref{ratios} shows two different behaviors of $R(t)$. 
For initial system states with an energy $\Op H_S(0)\gg\omega$ Fig. \ref{ratios} (a) shows $R(t)<1$, energy relaxation rate is faster than the dephasing rate (decreasing slope of the figure as a function of $t$).
We observe that for $\epsilon_0<\omega$ the ratio is
almost independent of $\omega_0$. In the weak coupling regime this region can be approximated by $R(t)\approx 1-\eta t/2$ \cite{Breuer}, dissipation 
happens twice faster than dephasing.
In contrast, Fig. \ref{ratios} (b) shows for an initial system state with $\Op H_S(0)\ll\omega$ that the dephasing rate is faster than the energy relaxation $R(t)>1$. Also the ratio is independent of $\omega_0$ for
frequencies $\epsilon_0<\omega$.
In both cases, when all spin frequencies are off-resonance with respect to 
the harmonic oscillator $\epsilon_0>\omega$, system and bath decouple freezing the dephasing $\Op a(t)=\Op a(0)$ and the energy exchange $\Op H_S(t)=\Op H_S(0)$, consequently $R(t)=1$.
It is possible to show that the dipolar interaction given by Eq. (\ref{relax}) can not produce pure dephasing due to the separation of time scales between system and bath. A necessary condition for pure dephasing is
that $\langle\Op H_S(t\rangle)$ is constant. Neglecting $\Op G(t)$ and $\Op G^{\dag}(t)$ in Eq. (\ref{Es}) we see that $\langle\Op H_S(t\rangle)$ is proportional to $|U_{00}(t)|^2$. From Eq. (\ref{Us}) we find
\beq
|U_{00}(t)|^2=\sum_{j=0}^K|y^{(j)}_0|^4+\sum_{j=0}^{K-1}\sum_{i>j}^{K}2\cos[t(\lambda_j-\lambda_i)]|y^{(j)}_0|^2|y^{(i)}_0|^2
\eeq
which remains constant if $\lambda_i=\lambda_j$ $\forall i$. From Eq. (\ref{lam}) it occurs if $\omega=\omega_i$ $\forall i$, all bath frequencies must be on-resonance with the system frequency. On 
the other hand, if all bath frequencies are equal, the coupling constants of the original system are $d_i=\sqrt{(J(\epsilon_i)/\rho(\epsilon_i))}\approx 0$ 
because $\rho(\epsilon_i)^{-1}\approx(\epsilon_{i+1}-\epsilon_i)\approx 0$. System and bath decouple due to the suppression of the interaction and no dephasing is produced. 
\section{Dephasing in the NV center}
\label{nv}
Once we presented different dephasing processes in a generic harmonic oscillator coupled to a spin bath, let us now consider a more realistic system.
We consider decoherence of a single NV center in diamond where the system, bath, and interaction
hamiltonians are very well known and their parameters can be easily controlled experimentally. 
The primary system is the NV center immersed in a bath caused by the electron spins of the surrounding nitrogen atoms. The NV center
is considered as a spin ($S_0=1$) placed in a static magnetic field $B$ along the $z$-axis. Its hamiltonian is \cite{hanson}
\beq
\Op H_S=D(\Op S_0^z)^2+g_0\mu_BB\Op S_0^z,
\label{hsnv}
\eeq
where $D=2.87$ GHz is the splitting between the levels $|m_S=0\rangle$ and $|m_S=\pm1\rangle$ at  $B=0$, $g_0=2$ is the Land\'e factor of the NV center and
$\mu_B$ is the Bohr's magneton. The first term of $\Op H_S$ represents the single-axis anisotropy whereas the second represents the Zeeman term due to the interaction
of the spin with the static magnetic field $B$. We are neglecting the hyperfine coupling, the interaction between the central spin ($S_0=1$) and the nuclear
spin $^{14}$N of the NV center.

The bath is constituted by nitrogen atoms each having an unpaired electron ($s_k=1/2$) interacting with the external magnetic field and with the rest of the spins through dipole-dipole
interactions. Again the nuclear spins of the nitrogens in the bath are not considered. The bath hamiltonian is given by \cite{hanson}
\beq
\Op H_B=g\mu_BB\sum_k^K\Op s_k^z+\frac{\mu_0\mu_B^2g^2}{4\pi}\sum_{j<k}^K\frac{1}{r^3_{jk}}[\mathbf{\Op s}_j\mathbf{\Op s}_k-3(\mathbf{\Op s}_j\mathbf{n}_{jk})(\mathbf{\Op s}_k\mathbf{n}_{jk})],
\label{hbnv}
\eeq
where $K$ is the total number of spins in the bath, $\mu_0$ the vacuum permeability, $g=2$ the Land\'e factor of the nitrogens, 
$\mathbf{\Op s}_k=(\Op s_k^x,\Op s_k^y,\Op s_k^z)=(\Op\sigma_k^x,\Op\sigma_k^y,\Op\sigma_k^z)/2$, and $\mathbf{r}_{jk}=(r^x_{jk},r^y_{jk},r^z_{jk})$ is the
relative position between spins $j$ and $k$ ($r_{jk}=|\mathbf{r}_{jk}|$ and $\mathbf{n}_{jk}=\mathbf{r}_{jk}/r_{jk}$).

The system-bath coupling is due to dipole-dipole interactions. Assuming that the NV center is placed at the origin \cite{hanson},
\beq
\Op H_{SB}=\frac{\mu_0\mu_B^2g_0g}{4\pi}\sum_k^K\frac{1}{r^3_{k}}[\mathbf{\Op S}_0\mathbf{\Op s}_k-3(\mathbf{\Op S}_0\mathbf{n}_{k})(\mathbf{\Op s}_k\mathbf{n}_{k})],
\label{hsbnv}
\eeq
with $\mathbf{\Op S}_0=(\Op S_0^x,\Op S_0^y,\Op S_0^z)$ and $\mathbf{r}_{k}=(r^x_{k},r^y_{k},r^z_{k})$ is the
relative position of spin $k$ with respect to the NV center ($r_{k}=|\mathbf{r}_{k}|$ and $\mathbf{n}_{k}=\mathbf{r}_{k}/r_{k}$).

The total hamiltonian $\Op H=\Op H_{S}+\Op H_{B}+\Op H_{SB}$ contains two different time scales, the anisotropy that produces the splitting between the $|m_S=0\rangle$ and $|m_S=\pm1\rangle$
levels set fast dynamics in the $ns$ regime whereas,  
typically, the dipole interaction strengths between electron spins $\gamma_{jk}=\mu_0\mu_B^2g^2/(4\pi r_{jk}^3)$ and $\gamma_k=\mu_0\mu_B^2gg_0/(4\pi r_{k}^3)$ are the order of MHz \cite{hanson} which
set a slower dynamics in the $\mu s$ regime.
\subsection{Standard pure dephasing}
Let us analyze the conditions where the total hamiltonian $\Op H=\Op H_{S}+\Op H_{B}+\Op H_{SB}$ given by Eqs. (\ref{hsnv}), (\ref{hbnv}), and (\ref{hsbnv}) produce 
standard pure dephasing. A strong magnetic field $B$ can be applied splitting the $|m_S = \pm1\rangle$ levels. As a result, the central NV spin can be formally treated as a 
pseudo-spin $\mathcal{S}_0=1/2$ connecting the $|m_S = 0\rangle$ and $|m_S = -1\rangle$ states \cite{hanson}, then the system hamiltonian takes the form,
\beq
\Op{\mathcal{H}}_S=D(\Op{\mathcal{S}}_0^z-1/2)^2+g_0\mu_BB (\Op{\mathcal{S}}_0^z-1/2).
\label{hsnvS}
\eeq
For the bath hamiltonian, if the magnetic field is much larger than the spin-spin coupling $g\mu_BB\gg \gamma_{jk}$, the rotating wave approximation can be
applied to Eq. (\ref{hbnv}) 
simplifying to an effective XXZ Ising-Heissenberg model \cite{nat_alex}, but because of the strong magnetic field applied along the $z$ axis the nitrogen spins
are preferably oriented in this direction, $\Op s_k^x$ and $\Op s_k^y\ll\Op s_k^z$. As a result the bath is governed by
 %
%
 %
\beq
\Op{\mathcal{H}}_{B}=g\mu_BB\sum_k^K\Op s_k^z+\sum_{j<k}^K\gamma_{jk}[1-3(r_{jk}^z)^2]\Op s_j^z\Op s_k^z,
\label{hbnvS2}
\eeq
If the magnetic field $B$ is also greater than the system-bath dipolar interaction $g\mu_BB\gg \gamma_k$
the pseudo-spin $\Op{\mathcal{S}_0}$ is also aligned with the magnetic field, consequently $\Op{\mathcal{S}}_0^x=\Op{\mathcal{S}}_0^y\approx 0$ and Eq. (\ref{hsbnv}) simplifies \cite{hanson},
\beq
\Op{\mathcal{H}}_{SB}=\sum_k^K\gamma_k[1-3(r_k^z)^2](\Op{\mathcal{S}}_0^z-1/2)\Op{s}_k^z.
\label{hbnvSB}
\eeq
Note that in the strong magnetic field regime $[\Op{\mathcal{H}}_{SB},\Op{\mathcal{H}}_S]=0$, thus no dissipation of the NV center takes place. Moreover as
the total hamiltonian $\Op{\mathcal{H}}=\Op{\mathcal{H}}_{S}+\Op{\mathcal{H}}_{B}+\Op{\mathcal{H}}_{SB}$ 
commutes with the pseudo-spin operator $[\Op{\mathcal{H}},\Op{\mathcal{S}}_0^z]=0$ the populations of the $|m_S=0\rangle$ and $|m_S=-1\rangle$ states do not evolve in time.
The NV center loses its coherence due to pure dephasing mechanism without exchanging energy with the bath. Finally, note that $[\Op{\mathcal{H}}_{SB},\Op{\mathcal{H}}_B]=0$ thus
the interaction does not change the bath excitations and the NV center experiences dephasing if initially the bath is activated.
%
%
%
%
%
%
%
%
%
%
%
%
%
%
%
%
%
%
\begin{figure}[t]
\begin{center}
\includegraphics[width=0.49\linewidth]{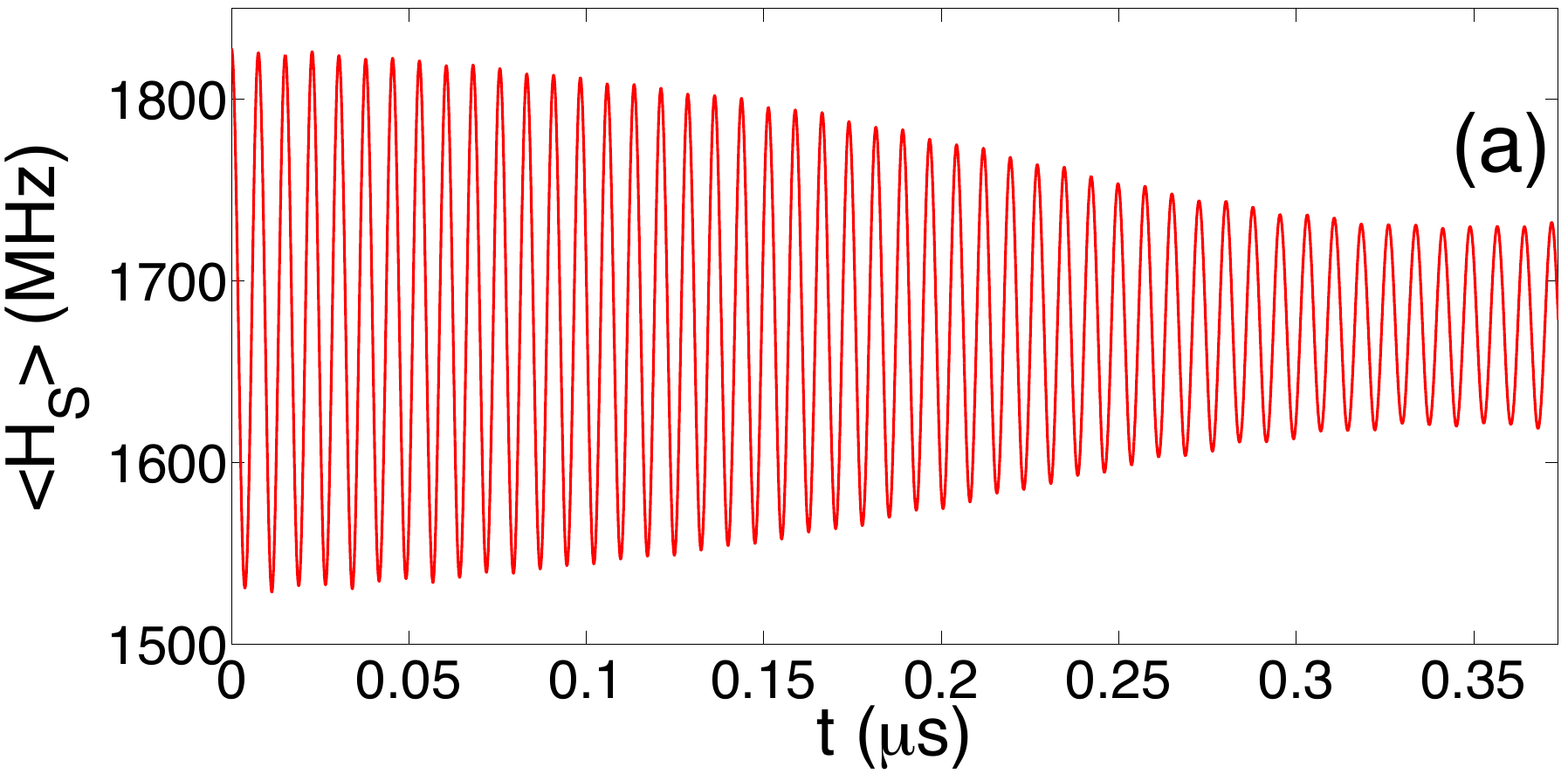}
\includegraphics[width=0.49\linewidth]{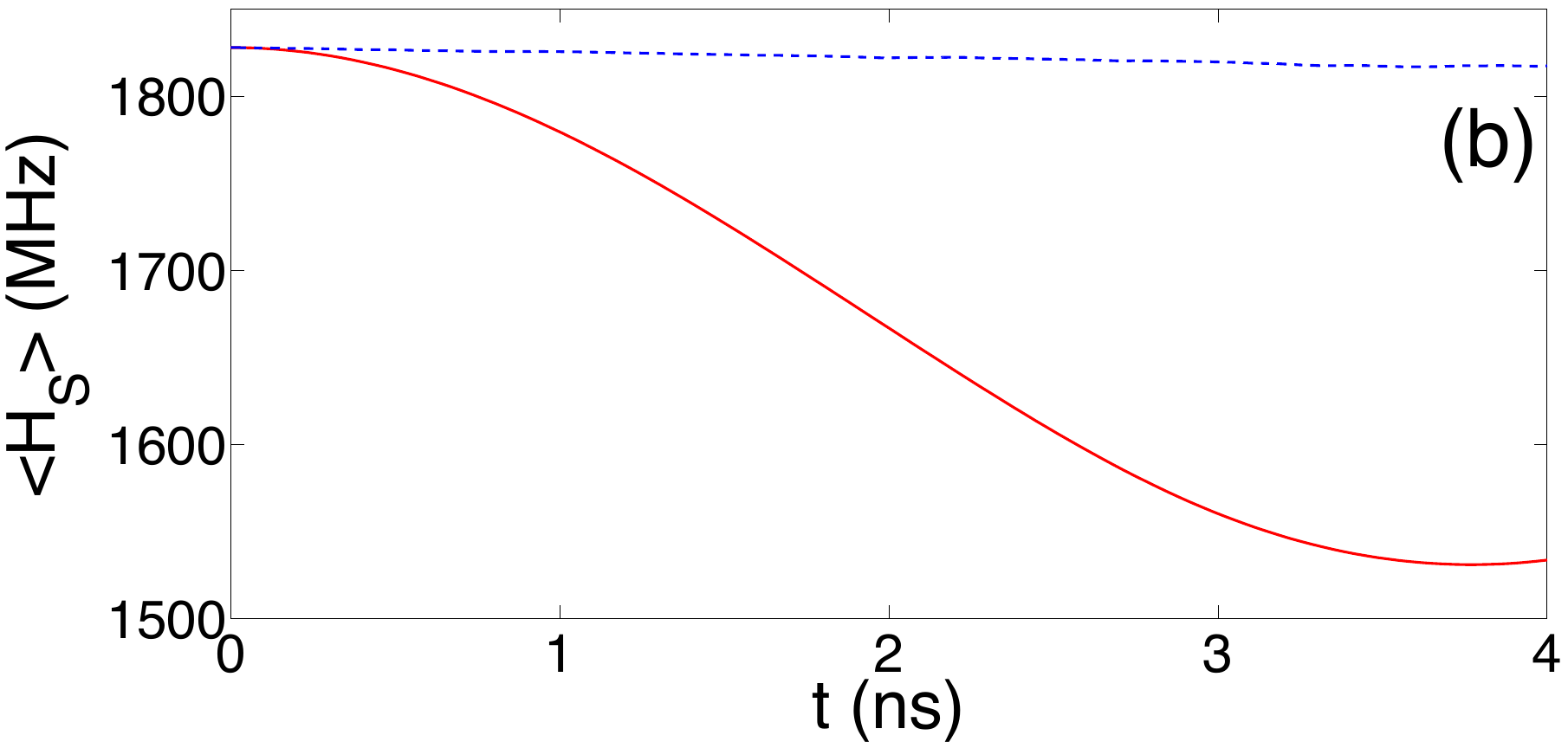}
\includegraphics[width=0.49\linewidth]{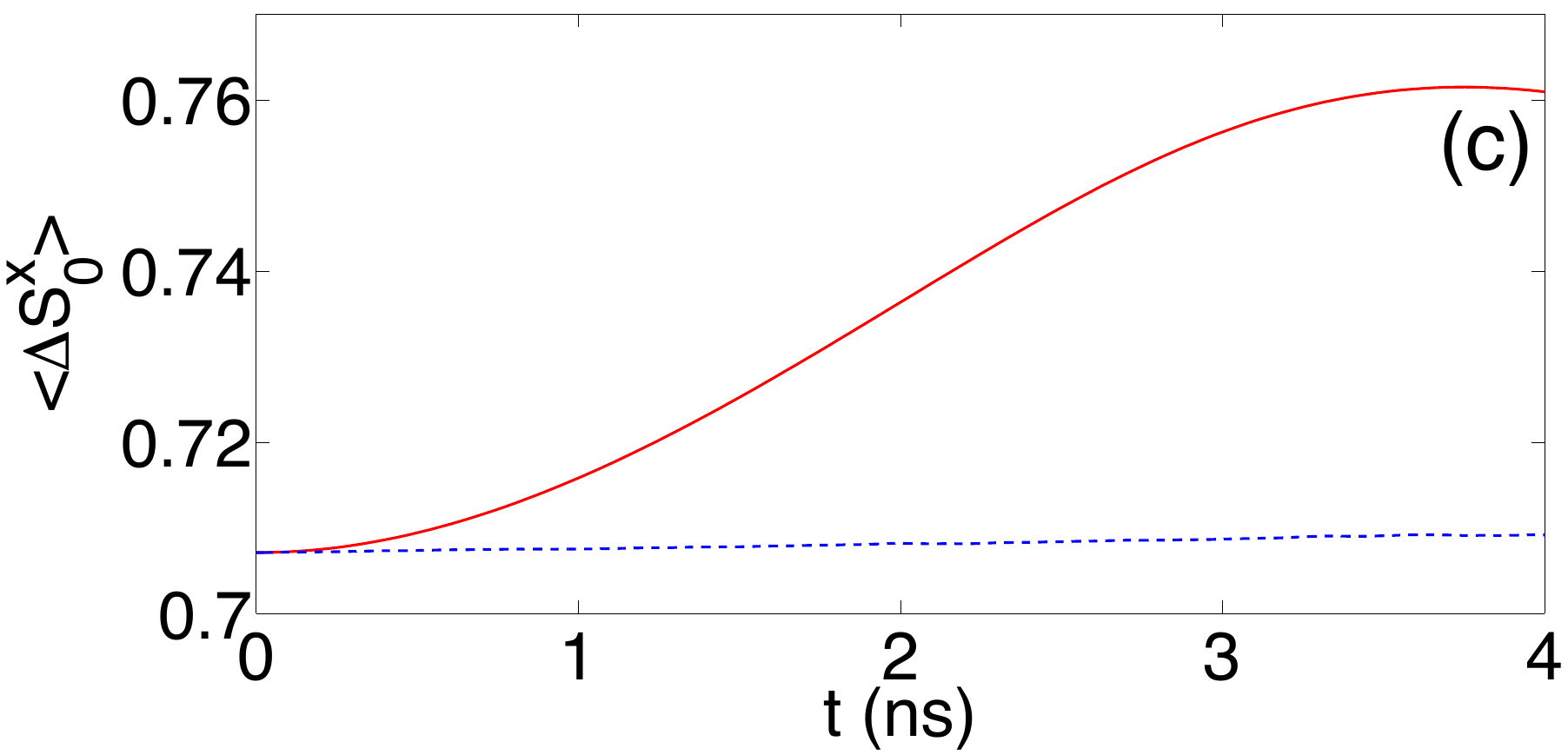}
\includegraphics[width=0.49\linewidth]{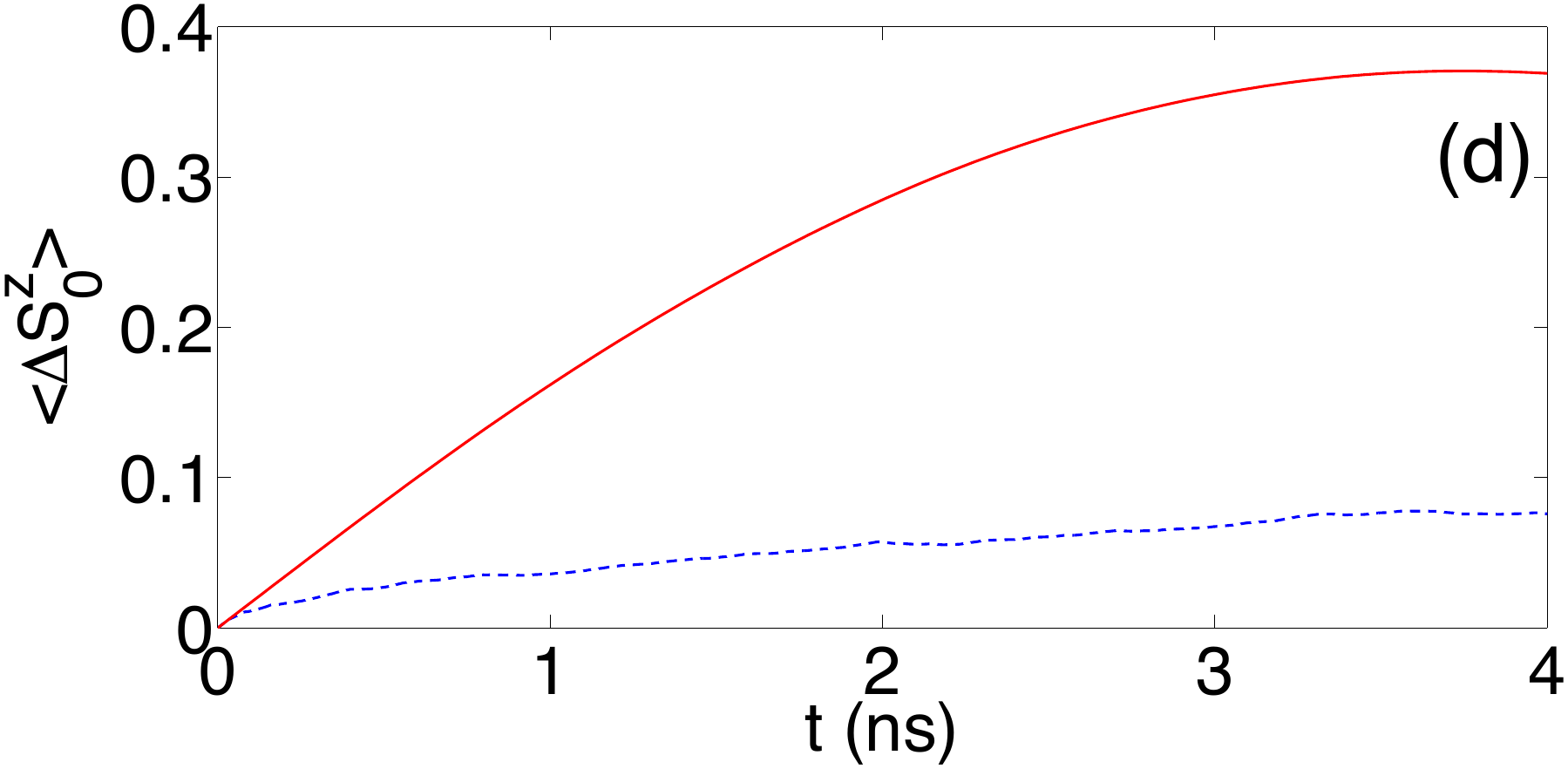}

\caption{(a) System energy as a function of time for the NV center. The fast oscillation in the $ns$ scale is produced by the anisotropy producing the  
$|m_S=0\rangle$ and $|m_S=\pm1\rangle$ level splitting. The energy decay in the $\mu s$ regime is produced by the dipolar interaction between system and bath.
(b) System energy as a function of time for the NV center in the $ns$ regime. The red solid line is the same as in (a). Swapping the spins of the bath 
at a rate $t_f\ll t_Z$ the system does not decay and remains in the excited state $|m_S=-1\rangle$ during the evolution without exchanging energy with the bath
(blue-dashed line). (Bottom) Standard deviation of the NV center spin in the $x$ direction (c) and $z$ direction (d) as a function of time. 
In the free evolution without spin swaps in the bath these functions oscillate in the $ns$ regime (red solid line). For spin swaps faster than $t_Z$ the fast oscillations
are suppressed and dephasing drives the NV center (blue-dashed line). 
Parameter values: $K=7$, $N_r=50$, $B=59$ $G$, and $\gamma_{jk}\approx \gamma_k\sim MHz$.
}
\label{NV}
\end{center}
\end{figure}
%
%
%
%
%
%
%
\subsection{Spin swap dephasing}
To simulate the NV center dynamics we assume that the total hamiltonian of the
whole system is the sum of Eqs. (\ref{hsnv}), (\ref{hbnv}) and (\ref{hsbnv}), where the nitrogen atoms that represent the bath are randomly distributed. 
The initial state $|\Psi(0)\rangle$ is prepared in such a way that the system, Eq. (\ref{hsnv}), is in the first excited state $|m_S=-1\rangle$ and the bath is initially 
completely de-excited (in the bit representation only the first spinor component is non-zero). As in the case of the harmonic oscillator, we compute the
dynamics of  $|\Psi(0)\rangle$ under the evolution of the total hamiltonian in the presence of spin swaps faster than $t_Z$ and in their absence, the results are plotted in Fig. \ref{NV}. 
When no spin swaps are produced, system and bath exchange energy on a time scale of $ns$
produced by the single-axis anisotropy as we infer from Fig. \ref{NV} (a). The dipole-dipole interaction between system and bath
makes the NV decay from the $|m_S=-1\rangle$ excited state at a $ms$ scale. Analyzing the decoherence through the standard deviation of the NV spin in different directions, we
observe a similar result, red solid lines in Figs. \ref{NV} (c) and (d).  
Both $\langle\Delta\Op S_0^x\rangle$ and $\langle\Delta\Op S_0^z\rangle$ show fast oscillations with a period of $ns$ eventually 
growing for a $ms$ time evolution. 

When spin swaps are performed in the bath with a rate faster than the Zeno time the situation changes drastically, dashed-blue lines of Fig. \ref{NV}. Looking at $\langle\Op H_S\rangle$ we observe in Fig. \ref{NV} (b) that
the NV center does not exchange energy with the bath remaining in the $|m_S=-1\rangle$
state. The system remains excited but not stationary, $\langle\Delta\Op S_0^x\rangle$ and $\langle\Delta\Op S_0^z\rangle$ are not constant and evolve slowly in time, Figs. \ref{NV} (c) and (d). 
The unitary evolution is suppressed by the Zeno effect and dephasing guides the system. Note that for this mechanism no initial activation of the bath was necessary.

%
%
%
%
%
%
%
%
%
%
%
%
%
%
%
\section{Discussion}
\label{discusion}
We have analyzed decoherence processes in open quantum systems produced by the interaction between a primary system and the surrounding environment. In an initial generic model constituted by
an harmonic oscillator coupled to a TLS bath we have presented a new pure dephasing mechanism which does not require initial activation of the bath and consequently is
temperature independent. Spin swap dephasing can be present even at low temperatures. 
In contrast, standard dephasing occurs when initially the bath is excited. Reducing the temperature, the initial number of excited bath modes, this mechanism can
be frozen out. 

For weak coupling
we show that the dipolar interaction can not produce pure dephasing due to the separation of time scales between the system and the bath. However, dephasing can occur but producing also
energy dissipation. The ratio between dissipation and dephasing depends on the initial state energy.

Activated and non activated dephasing processes have been demonstrated for a NV center where the primary system, interaction, and bath can be accurately modified applying a magnetic field
and controlling the density of impurities. In the activated process a strong magnetic field aligns the NV and nitrogen spins. As result, the interaction produces the dephasing of the primary system without 
energy exchange with the bath. This process occurs if initially the bath is excited and consequently is temperature dependent. Non activated dephasing is also demonstrated in the NV center,
swapping nitrogen atoms at a rate faster than the characteristic Zeno time of the system.

We have presented here a new dephasing mechanism in order to clarify the relation between dephasing and the initial bath configuration, but there remain interesting other questions. For example, 
the study of the conditions for dephasing considering baths not only with single spin-spin interactions but higher order processes or analyzing dephasing
for non linear system-bath couplings.

\ack
We thank D. Gelman, A. Levy, and C. P. Koch for fruitful discussions. We acknowledge    
funding by the Israel Science Foundation, the Basque Government
(Grant No. IT472-10), Ministerio de 
Econom\'ia y Competitividad (FIS2012-36673- C03-01),  the program UFI 11/55 of UPV/EHU, and
COST Action MP1209 ``Thermodynamics in the quantum regime''.
E. T. is supported by the Basque Government postdoctoral program. \\
\appendix

\section{Explicit expression for the evolution operator}\label{U00}

It is possible to write the components of the evolution operator $\mathds{U}(t)=\mathds{A}e^{-i\lambda t}\mathds{A}^{\dag}$ in terms of the eigenvalues and eigenvectors of
the hamiltonian Eq. (\ref{Hho}). The characteristic equation $\det (\mathds{H} -\lambda\mathds{1})=0$ can be rewritten as \cite{braun}
\beq
\label{lam}
\omega-\lambda=\sum_{i=1}^K|\chi_i|^2/(\omega_i-\lambda)
\eeq
which is the transcendental equation that defines the eigenvalues of (\ref{Hho}). The eigenvectors are defined by
\beq
\mathds{H}y^{(i)}=\lambda_iy^{(i)} \quad (i=0,1,...,K),
\eeq
with the aid of Eq. (\ref{cHc}) the $j$th components of the $i$th eigenvector can be expressed in terms of the $y^{(i)}_0$ component,
\beq
y^{(i)}_j=-\frac{\chi_jy_0^{(i)}}{\omega_j-\lambda_i}\quad (i=0,1,...,K) \quad (j=1,2,...,K).
\eeq 
Using the orthonormalization condition for the eigenvectors $\sum_i|y^{(i)}|^2=1$ we find
\beq
|y_0^{(i)}|^2=\bigg(1+\sum_{j=1}^K\frac{|\chi_j|^2}{(\omega_j-\lambda_i)^2}\bigg)^{-1}
\eeq
and finally the transformation $\mathds{A}$ that diagonalizes the hamiltonian is explicitly written as
\beq
\mathds{A}=(y^{(0)},y^{(1)},...,y^{(K)})=
\left(\begin{array}{cccc}
y_0^{(0)} & y_0^{(1)} &...& y_0^{(K)}\\
\frac{y_0^{(0)}\chi_1}{\lambda_0-\omega_1} & \frac{y_0^{(1)}\chi_1}{\lambda_1-\omega_1} & ... & \frac{y_0^{(K)}\chi_1}{\lambda_K-\omega_1} \\
\vdots&\vdots&\ddots&\vdots\\
\frac{y_0^{(0)}\chi_K}{\lambda_0-\omega_K} & \frac{y_0^{(1)}\chi_K}{\lambda_1-\omega_K} & ... & \frac{y_0^{(K)}\chi_K}{\lambda_K-\omega_K} 
\end{array}\right).
\label{eigenval}
\eeq
The different components of the matrix corresponding to the evolution operator $\mathds{U}(t)=\mathds{A}e^{-i\lambda t}\mathds{A}^{\dag}$ are
\beqa
\label{Us}
U_{00}(t)&=&\sum_{j=0}^K |y_0^{(j)}|^2e^{-i\lambda_jt}\nonumber \\
U_{0k}(t)&=&U_{k0}(t)=\sum_{j=0}^K |y_0^{(j)}|^2\frac{\chi_k}{\lambda_j-\omega_k}e^{-i\lambda_jt} \quad (k\neq 0) \nonumber \\
U_{jk}(t)&=&\sum_{l=0}^K |y_0^{(l)}|^2 \frac{\chi_j\chi_k}{(\lambda_l-\omega_j)(\lambda_l-\omega_k)}e^{-i\lambda_lt} \quad (j,k\neq 0).
\eeqa

\end{document}